\documentclass[twocolumn]{aastex62}

\usepackage{verbatim}
\usepackage[usenames, dvipsnames]{}

\newcommand{\TESS}{\emph{TESS}}

\newcommand{\bjdtdb}{\ensuremath{\rm {BJD_{TDB}}}}

\newcommand{\fave}{\langle F \rangle}

\newcommand{\gcmc}{\ensuremath{\rm g\,cm^{-3}}}

\newcommand{\kms}{km\,s$^{-1}$}

\newcommand{\logg}{\ensuremath{\log g}}

\newcommand{\lsun}{\ensuremath{L_\sun}}

\newcommand{\mbol}{$m_{\rm bol}$}
\newcommand{\mearth}{\ensuremath{M_\earth}}
\newcommand{\me}{\ensuremath{\,M_{\rm E}}}

\newcommand{\mstar}{\ensuremath{M_\star}}
\newcommand{\msun}{\ensuremath{M_\sun}}
\newcommand{\ms}{\ensuremath{\rm m\,s^{-1}}}

\newcommand{\rearth}{\ensuremath{R_\earth}}
\newcommand{\re}{\ensuremath{\,R_{\rm E}}}

\newcommand{\rhostar}{\ensuremath{\rho_\star}}

\newcommand{\rstar}{\ensuremath{R_\star}}
\newcommand{\rsun}{\ensuremath{R_\sun}}

\newcommand{\Teff}{$T_{\rm eff}$}
\newcommand{\teff}{$T_{\rm eff}$}
\newcommand{\jwst}{\emph{JWST}}
\newcommand{\Lya}{Ly$\alpha$}

\newcommand{\thisstar}{TOI-1231}
\newcommand{\thisplanetb}{TOI-1231~b}
\newcommand{\plradunc}{3.65$^{+0.16}_{-0.15}$ \rearth} 

\newcommand{\plmassunc}{15.5$\pm$3.3 \mearth}
\newcommand{\plrhounc}{1.74$^{+0.47}_{-0.42}$ g\,cm$^{-3}$ } 

\shorttitle{A temperate Neptune around TOI-1231}
\shortauthors{Burt et al. 2021}

\begin{document}

\title{TOI-1231~b: A Temperate, Neptune-Sized Planet Transiting the Nearby M3 Dwarf NLTT 24399}

\author[0000-0002-0040-6815]{Jennifer~A.~Burt}
\affiliation{Jet Propulsion Laboratory, California Institute of Technology, 4800 Oak Grove Drive, Pasadena, CA 91109, USA}

\author[0000-0003-2313-467X]{Diana~Dragomir}
\affiliation{Department of Physics and Astronomy, University of New Mexico, 1919 Lomas Blvd NE, Albuquerque, NM 87131, USA}

\author{Paul Molli{\`e}re}
\affiliation{Max-Planck-Institut f\"{u}r Astronomie, K\"{o}nigstuhl 17, 69117, Heidelberg, Germany}

\author{Allison Youngblood}
\affiliation{Laboratory for Atmospheric and Space Physics, 1234 Innovation Drive, Boulder, CO, 80303, USA}

\author[0000-0003-1756-4825]{Antonio Garc\'{i}a Mu\~{n}oz}
\affiliation{AIM, CEA, CNRS, Universit\'e Paris-Saclay, Universit\'e de Paris,
Gif-sur-Yvette, France.}

\author{John McCann}
\affiliation{Department of Physics, University of California, Santa Barbara, CA 93106, USA}

\author[0000-0003-0514-1147]{Laura Kreidberg}
\affiliation{Max-Planck-Institut f\"{u}r Astronomie, K\"{o}nigstuhl 17, 69117, Heidelberg, Germany}

\author[0000-0003-0918-7484]{Chelsea~ X.~Huang}
\altaffiliation{Juan Carlos Torres Fellow}
\affiliation{Department of Physics and Kavli Institute for Astrophysics and Space Research, Massachusetts Institute of Technology, Cambridge, MA 02139, USA}

\author[0000-0001-6588-9574]{Karen~A.~Collins} 
\affiliation{Center for Astrophysics ${\rm \mid}$ Harvard {\rm \&} Smithsonian, 60 Garden Street, Cambridge, MA 02138, USA}

\author[0000-0003-3773-5142]{Jason~D.~Eastman}
\affiliation{Center for Astrophysics ${\rm \mid}$ Harvard {\rm \&} Smithsonian, 60 Garden Street, Cambridge, MA 02138, USA}

\author{Lyu Abe}
\affiliation{Universit\'e C\^ote d'Azur, Observatoire de la C\^ote d'Azur, CNRS, Laboratoire Lagrange, Bd de l'Observatoire, CS 34229, 06304 Nice cedex 4, France}

\author{Jose M. Almenara}
\affiliation{Univ. Grenoble Alpes, CNRS, IPAG, 38000, Grenoble, France}

\author{Ian~J.~M.~Crossfield}
\affiliation{Physics \& Astronomy Department, University of Kansas, Lawrence, KS, USA}

\author{Carl Ziegler}
\affiliation{Dunlap Institute for Astronomy and Astrophysics, University of Toronto, 50 St. George Street, Toronto, Ontario M5S 3H4, Canada}

\author[0000-0001-8812-0565]{Joseph~E.~Rodriguez}
\affiliation{Center for Astrophysics ${\rm \mid}$ Harvard {\rm \&} Smithsonian, 60 Garden Street, Cambridge, MA 02138, USA}
\affiliation{Department of Physics and Astronomy, Michigan State University, East Lansing, MI, 48824 USA}

\author[0000-0003-2008-1488]{Eric~E.~Mamajek}
\affiliation{Jet Propulsion Laboratory, California Institute of Technology, 4800 Oak Grove Drive, Pasadena, CA 91109, USA}

\author[0000-0002-3481-9052]{Keivan~G.~Stassun} 
\affiliation{Department of Physics and Astronomy, Vanderbilt University, Nashville, TN 37235, USA}
\affiliation{Department of Physics, Fisk University, Nashville, TN 37208, USA}

\author[0000-0003-1312-9391]{Samuel~P.~Halverson}
\affiliation{Jet Propulsion Laboratory, California Institute of Technology, 4800 Oak Grove Drive, Pasadena, CA 91109, USA}

\author[0000-0001-6213-8804]{Steven~Jr.~Villanueva}
\altaffiliation{Pappalardo Fellow}
\affiliation{Department of Physics and Kavli Institute for Astrophysics and Space Research, Massachusetts Institute of Technology, Cambridge, MA 02139, USA}

\author[0000-0003-1305-3761]{R.~Paul~Butler}
\affiliation{Earth \& Planets Laboratory, Carnegie Institution for Science, 5241 Broad Branch Road, NW, Washington, DC 20015, USA}

\author[0000-0002-6937-9034]{Sharon Xuesong Wang}
\affiliation{The Observatories of the Carnegie Institution for Science, 813 Santa Barbara Street, Pasadena, CA 91101, USA}
\affiliation{Department of Astronomy, Tsinghua University, Beijing 100084, People's Republic of China}

\author[0000-0001-8227-1020]{Richard P. Schwarz}
\affiliation{Patashnick Voorheesville Observatory, Voorheesville, NY 12186, USA}

\author[0000-0003-2058-6662]{George~R.~Ricker}
\affiliation{Department of Physics and Kavli Institute for Astrophysics and Space Research, Massachusetts Institute of Technology, Cambridge, MA 02139, USA}

\author[0000-0001-6763-6562]{Roland~Vanderspek}
\affiliation{Department of Physics and Kavli Institute for Astrophysics and Space Research, Massachusetts Institute of Technology, Cambridge, MA 02139, USA}

\author[0000-0001-9911-7388]{David~W.~Latham}
\affiliation{Center for Astrophysics ${\rm \mid}$ Harvard {\rm \&} Smithsonian, 60 Garden Street, Cambridge, MA 02138, USA}

\author[0000-0002-6892-6948]{S.~Seager}
\affiliation{Department of Physics and Kavli Institute for Astrophysics and Space Research, Massachusetts Institute of Technology, Cambridge, MA 02139, USA}
\affiliation{Department of Earth, Atmospheric and Planetary Sciences, Massachusetts Institute of Technology, Cambridge, MA 02139, USA}
\affiliation{Department of Aeronautics and Astronautics, MIT, 77 Massachusetts Avenue, Cambridge, MA 02139, USA}

\author[0000-0002-4265-047X]{Joshua~N.~Winn}
\affiliation{Department of Astrophysical Sciences, Princeton University, 4 Ivy Lane, Princeton, NJ 08544, USA}

\author[0000-0002-4715-9460]{Jon~M.~Jenkins}
\affiliation{NASA Ames Research Center, Moffett Field, CA 94035, USA}

\author{Abdelkrim Agabi}
\affiliation{Universit\'e C\^ote d'Azur, Observatoire de la C\^ote d'Azur, CNRS, Laboratoire Lagrange, Bd de l'Observatoire, CS 34229, 06304 Nice cedex 4, France}

\author[0000-0001-9003-8894]{Xavier Bonfils}
\affiliation{Univ. Grenoble Alpes, CNRS, IPAG, 38000, Grenoble, France}


\author{David Ciardi}
\affiliation{NASA Exoplanet Science Institute, Infrared Processing \& Analysis Center, Jet Propulsion Laboratory, California Institute of Technology, Pasadena CA 91125, USA}

\author{Marion Cointepas}
\affiliation{Univ. Grenoble Alpes, CNRS, IPAG, 38000, Grenoble, France}
\affiliation{Observatoire de l'Universit\'e de Gen\'eve, Chemin des Maillettes 51, 1290 Versoix, Switzerland}

\author[0000-0002-5226-787X]{Jeffrey~D.~Crane}
\affiliation{The Observatories of the Carnegie Institution for Science, 813 Santa Barbara Street, Pasadena, CA 91101, USA}

\author{Nicolas Crouzet}
\affiliation{European Space Agency (ESA), European Space Research and Technology Centre (ESTEC), Keplerlaan 1, 2201 AZ Noordwijk, The Netherlands}

\author[0000-0002-3937-630X]{Georgina Dransfield}
\affiliation{School of Physics \& Astronomy, University of Birmingham, Edgbaston, Birmingham, B15 2TT, UK}

\author[0000-0001-6039-0555]{Fabo Feng}
\affiliation{Tsung-Dao Lee Institute, Shanghai Jiao Tong University, 800 Dongchuan Road, Shanghai 200240, People's Republic of China}
\affiliation{Department of Astronomy, School of Physics and Astronomy, Shanghai Jiao Tong University, 800 Dongchuan Road, Shanghai 200240, People's Republic of China}

\author{Elise Furlan}
\affiliation{NASA Exoplanet Science Institute, Infrared Processing \& Analysis Center, Jet Propulsion Laboratory, California Institute of Technology, Pasadena CA 91125, USA}

\author[0000-0002-7188-8428]{Tristan Guillot} 
\affiliation{Universit\'e C\^ote d'Azur, Observatoire de la C\^ote d'Azur, CNRS, Laboratoire Lagrange, Bd de l'Observatoire, CS 34229, 06304 Nice cedex 4, France}

\author[0000-0002-5463-9980]{Arvind F. Gupta}
\affiliation{Department of Astronomy \& Astrophysics, 525 Davey Laboratory, The Pennsylvania State University, University Park, PA, 16802, USA}
\affiliation{Center for Exoplanets and Habitable Worlds, 525 Davey Laboratory, The Pennsylvania State University, University Park, PA, 16802, USA}

\author[0000-0002-2532-2853]{Steve B. Howell}
\affiliation{NASA Ames Research Center, Moffett Field, CA 94035, USA}

\author[0000-0002-4625-7333]{Eric L. N. Jensen}
\affiliation{Dept.\ of Physics \& Astronomy, Swarthmore College, Swarthmore PA 19081, USA}

\author{Nicholas Law}
\affiliation{Department of Physics and Astronomy, The University of North Carolina at Chapel Hill, Chapel Hill, NC 27599-3255, USA}

\author[0000-0003-3654-1602]{Andrew W. Mann}
\affiliation{Department of Physics and Astronomy, The University of North Carolina at Chapel Hill, Chapel Hill, NC 27599-3255, USA}

\author{Wenceslas Marie-Sainte}
\affiliation{Concordia Station, IPEV/PNRA, Antarctica}

\author[0000-0001-7233-7508]{Rachel A.~Matson}
\affiliation{U.S.~Naval Observatory, Washington, DC 20392, USA}

\author{Elisabeth~C.~Matthews}
\affil{Observatoire de l’Universit\'e de Gen\`eve, Chemin des Maillettes 51, 1290 Versoix, Switzerland}

\author[0000-0001-5000-7292]{Djamel M\'ekarnia}
\affiliation{Universit\'e C\^ote d'Azur, Observatoire de la C\^ote d'Azur, CNRS, Laboratoire Lagrange, Bd de l'Observatoire, CS 34229, 06304 Nice cedex 4, France}

\author[0000-0002-3827-8417]{Joshua Pepper}
\affiliation{Department of Physics, Lehigh University, 16 Memorial Drive East, Bethlehem, PA 18015, USA}

\author{Nic Scott}
\affiliation{NASA Ames Research Center, Moffett Field, CA 94035, USA}

\author{Stephen~A.~Shectman}
\affiliation{The Observatories of the Carnegie Institution for Science, 813 Santa Barbara Street, Pasadena, CA 91101, USA}

\author[0000-0001-5347-7062]{Joshua E. Schlieder}
\affiliation{Exoplanets and Stellar Astrophysics Laboratory, Code 667, NASA Goddard Space Flight Center, Greenbelt, MD, 20771, USA}

\author{Fran\c{c}ois-Xavier Schmider}
\affiliation{Universit\'e C\^ote d'Azur, Observatoire de la C\^ote d'Azur, CNRS, Laboratoire Lagrange, Bd de l'Observatoire, CS 34229, 06304 Nice cedex 4, France}

\author[0000-0002-5951-8328]{Daniel J. Stevens}
\altaffiliation{Eberly Research Fellow}
\affiliation{\psu}
\affiliation{\psust}

\newcommand{\psu}{Department of Astronomy \& Astrophysics, The Pennsylvania State University, 525 Davey Lab, University Park, PA 16802, USA}
\newcommand{\psust}{Center for Exoplanets and Habitable Worlds, The Pennsylvania State University, 525 Davey Lab, University Park, PA 16802, USA}

\author{Johanna~K.~Teske}
\affiliation{Earth \& Planets Laboratory, Carnegie Institution for Science, 5241 Broad Branch Road, NW, Washington, DC 20015, USA}

\author{Amaury H.M.J. Triaud}
\affiliation{School of Physics \& Astronomy, University of Birmingham, Edgbaston, Birmingham, B15 2TT, UK}

\author[0000-0002-9003-484X]{David~Charbonneau}
\affiliation{Center for Astrophysics ${\rm \mid}$ Harvard {\rm \&} Smithsonian, 60 Garden Street, Cambridge, MA 02138, USA}

\author[0000-0002-3321-4924]{Zachory~K.~Berta-Thompson}
\affiliation{University of Colorado Boulder, Boulder, CO 80309, USA}

\author[0000-0002-7754-9486]{Christopher~J.~Burke}
\affiliation{Department of Physics and Kavli Institute for Astrophysics and Space Research, Massachusetts Institute of Technology, Cambridge, MA 02139, USA}

\author[0000-0002-6939-9211]{Tansu~Daylan}
\altaffiliation{Kavli Fellow}
\affiliation{Department of Physics and Kavli Institute for Astrophysics and Space Research, Massachusetts Institute of Technology, Cambridge, MA 02139, USA}

\author[0000-0001-7139-2724]{Thomas~Barclay}
\affiliation{NASA Goddard Space Flight Center, 8800 Greenbelt Road, Greenbelt, MD 20771, USA}
\affiliation{University of Maryland, Baltimore County, 1000 Hilltop Circle, Baltimore, MD 21250, USA}

\author[0000-0002-5402-9613]{Bill Wohler}
\affiliation{SETI Institute, Mountain View, CA 94043, USA} 
\affiliation{NASA Ames Research Center, Moffett Field, CA 94035, USA}

\author[0000-0002-9314-960X]{C. E. Brasseur}
\affiliation{Space Telescope Science Institute}

\begin{abstract}
We report the discovery of a transiting, temperate, Neptune-sized exoplanet orbiting the nearby ($d$ = 27.5 pc), M3V star \object{TOI-1231} (\object{NLTT~24399}, \object{L~248-27}, \object{2MASS~J10265947-5228099}). The planet was detected using photometric data from the \textit{Transiting Exoplanet Survey Satellite} and followed up with observations from the Las Cumbres Observatory and the Antarctica Search for Transiting ExoPlanets program. Combining the photometric data sets, we find that the newly discovered planet has a radius of \plradunc, and an orbital period of 24.246 days. Radial velocity measurements obtained with the Planet Finder Spectrograph on the Magellan Clay telescope confirm the existence of the planet and lead to a mass measurement of \plmassunc. With an equilibrium temperature of just 330K TOI-1231~b is one of the coolest small planets accessible for atmospheric studies thus far, and its host star's bright NIR brightness (J=8.88, K$_{s}$=8.07) make it an exciting target for HST and JWST. Future atmospheric observations would enable the first comparative planetology efforts in the 250-350 K temperature regime via comparisons with K2-18~b. Furthermore, TOI-1231's high systemic radial velocity (70.5 k\ms) may allow for the detection of low-velocity hydrogen atoms escaping the planet by Doppler shifting the H I Ly-alpha stellar emission away from the geocoronal and ISM absorption features.
\end{abstract}


\section{Introduction} \label{sec:intro}


The observing strategy adopted by NASA's Transiting Exoplanet Survey Satellite \citep[\TESS,][]{Ricker2014}, wherein each hemisphere is divided into 13 sectors each of which is surveyed for roughly 28 days, is producing the most comprehensive all-sky search for transiting planets. This approach has already proven its capability to detect both large and small planets \citep{Wang2019, Rodriguez2019, Dragomir2019, Luque2019, Burt2020} around stars ranging from Sun-like \citep{Huang2018} down to low-mass M dwarf stars \citep{Vanderspek2019}. 

Although it enables the detection of exoplanets across the sky, TESS's survey strategy also produces significant observational biases based on orbital period. Exoplanets must transit their host stars at least twice within \TESS's observing span in order to be detected with the correct period by the Science Processing Operations Center (SPOC) pipeline, which searches the 2-minute cadence TESS data obtained for pre-selected target stars \citep{Jenkins2016}. Because 74\% of \TESS' total sky coverage is only observed for $\leq$28 days, the majority of \TESS\ exoplanets detected by the SPOC are expected to have periods less than 14 days. Simulations of the \TESS\ exoplanet yield presented in \citet{Sullivan2015} find mean and median orbital periods of 13.48 and 8.19 days, respectively, among all detected planets. Similarly, \citet{Barclay18} found mean and median periods of 10.23 and 7.03 days among the full set of detected planets, and those values drop to just 7.42 and 5.89 days when considering the stars only observed during a single \TESS\ sector. And additional simulations from \citet{Jiang2019} show that for stars observed only in a single sector the expected mean value of the most frequently detected orbital period is 5.01 days, with a most detected range of 2.12 to 11.82 days. Even when considering the Ecliptic poles, where \TESS\ carries out 351 days of observing coverage during its primary mission, the expected mean orbital period is still only 10.93 days with a most detected range from 3.35 to 35.65 days. 

Of the 1994 \TESS\ objects of interest (TOIs) identified as planet candidates using Sectors 1-26 of the primary mission, only $\sim$14\% have periods longer than 14 days\footnote{https://tev.mit.edu/data/collection/193/}. But these longer period (and thereby cooler) candidates are some of the most intriguing targets for atmospheric characterization. This is especially true for Neptune sized planets whose lower temperatures could spark several marked changes in the expected atmospheric chemistry: disequilibrium due to rain out is relevant, water clouds may form, and ammonia is the dominant carrier of nitrogen \citep[see, e.g.][]{Morley2014}. Thus these cooler TOIs merit additional attention and focused follow up efforts both to confirm their planetary nature and to obtain the precise mass measurements necessary for correct interpretation of future transmission spectroscopy observations \citep{Batalha2019}.

Here we report the discovery of a Neptune-sized planet
transiting \object{TOI-1231} (NLTT 24399, 
L~248-27, TIC~447061717, 
2MASS~J10265947-5228099), a $V$ = 12.3 mag M3V star \citep{Gaidos2014} in the Vela constellation at $d$ = 27.493\,$\pm$\,0.0123 pc 
\citep[$\varpi$\,=\,36.3726\,$\pm$\,0.0163 mas; GaiaEDR3][]{GaiaEDR3}. This paper is organized as follows. In Section 2 we characterize the host star using details from published catalogs and new data obtained once the \TESS\ planet candidate was identified. In Section 3 we describe the initial discovery of TOI-1231 b and the follow up data obtained in an effort to characterize the planet. In Section 4 we outline the procedure used to perform a joint fit to the host star and planet, and then in Section 5 we conclude with a discussion of the planet's promising potential for future atmospheric characterization. 


\section{Stellar Data \& Characterization \label{sec:stellardata}}

\subsection{Background}

TOI-1231 was first reported as a high proper motion star
(0\farcs 35 yr$^{-1}$) by \citet[][as LTT 3840 and 
L 248-27]{Luyten1957}, and later in the Revised NLTT catalog 
\citep[][as NLTT 24399]{Luyten1979}.
Over the past decade, the star has appeared in several surveys 
of high proper motion 2MASS and WISE stars and nearby M dwarfs 
\citep[e.g.][]{Lepine2011, Frith2013, Kirkpatrick2014, 
Schneider2016}.
The only previous spectral characterization of TOI-1231
was by \citet{Gaidos2014} in the CONCH-SHELL survey,
who reported a spectral type of M3V.
Previous estimates of the basic stellar parameters
were reported by \citet{Gaidos2014, Muirhead2018, Stassun2019}.

\begin{table}[htbp]
    \caption{Astrometry \& Photometry for TOI-1231}\label{tab:star}
    \begin{tabular}{lcc}
    \hline\hline
    Parameter & Value & Source \\
     \hline \hline
    Designations & TIC 447061717  & \citet{Stassun2019}\\
                 & NLTT 24399 & \citet{Luyten1979}\\
    RA (ICRS, J2000) & 10:26:59.34 & Gaia EDR3\\
    Dec (ICRS, J2000) & -52 28 04.16 & Gaia EDR3\\
    $\mu$ RA (mas yr$^{-1}$) & -89.394 $\pm$ 0.019 & Gaia EDR3\\
    $\mu$ Dec (mas yr$^{-1}$)& 361.546 $\pm$ 0.015 & Gaia EDR3\\
    Parallax (mas) & 36.3726 $\pm$ 0.0163 & Gaia EDR3$^a$\\
    Distance (pc) & 27.4932 $\pm$ 0.0123 & Gaia EDR3$^a$\\
    $v_R$ (\kms) &  70.48\,$\pm$\,0.54 & Gaia DR2\\
    SpT & M3V & \citet{Gaidos2014}\\
    \hline
     $B$  & 13.739 $\pm$ 0.028 & APASS/DR10\\
     $V$  & 12.322 $\pm$ 0.023 & APASS/DR10\\
     $g'$ & 12.997 $\pm$ 0.044 & APASS/DR10\\
     $r'$ & 11.806 $\pm$ 0.070 & APASS/DR10\\
     $i'$ & 10.754 $\pm$ 0.175 & APASS/DR10\\
     {\it TESS} & 10.2565 $\pm$ 0.007 & TIC8\\
      $G$      & 11.3612 $\pm$ 0.0009 & Gaia EDR3\\
      $G_{BP}$ & 12.5440 $\pm$ 0.0022 & Gaia EDR3\\
      $G_{RP}$ & 10.2735 $\pm$ 0.0015 & Gaia EDR3\\
      $J$      & 8.876 $\pm$ 0.027 & 2MASS\\
      $H$      & 8.285 $\pm$ 0.038 & 2MASS\\
      $K_{s}$  & 8.069 $\pm$ 0.026 & 2MASS\\
      $W_{1}$  & 7.922 $\pm$ 0.024 & WISE\\
      $W_{2}$  & 7.826 $\pm$ 0.020 & WISE\\
      $W_{3}$  & 7.732 $\pm$ 0.017 & WISE\\
      $W_{4}$  & 7.515 $\pm$ 0.091 & WISE\\
\hline
    U (km s$^{-1}$) & -20.25\,$\pm$\,0.24 & This work\\ 
    V (km s$^{-1}$) & -73.37\,$\pm$\,0.36 & This work\\
    W (km s$^{-1}$) &  39.45\,$\pm$\,0.33 & This work\\  
    S$_{tot}$ (km s$^{-1}$) & 85.73\,$\pm$\,0.35 & This work\\
\hline
\hline
\multicolumn{3}{l}{$^a$: Correction of -17 $\mu$as applied to the Gaia parallax following}\\ 
\multicolumn{3}{l}{the prescription in \citet{Lindegren2020}} \\
    \end{tabular}
\end{table}

\subsection{Astrometry \& Photometry}

Stellar astrometry and visible and infrared photometry are compiled in Table \ref{tab:star}. 
The positions, proper motions, parallax, and {\it Gaia} photometry are from Gaia EDR3 \citep{GaiaEDR3}, while the radial velocity is from GaiaDR2 \citep{GaiaDR2}.
We convert the astrometry to Galactic velocities following 
\citep{ESA1997}\footnote{$U$ towards Galactic center, $V$ in direction of Galactic spin, and $W$ towards North Galactic Pole \citep{ESA1997}.}.
Photometry is reported from APASS Data Release 10 \citep{Henden2016}\footnote{https://www.aavso.org/apass}, the TESS Input Catalog (TIC8), 2MASS \citep{Cutri2003}, and WISE \citep{Cutri2012}.

\subsection{Spectral Energy Distribution \label{sec:SED}}

\begin{figure}
\includegraphics[width=.45 \textwidth]{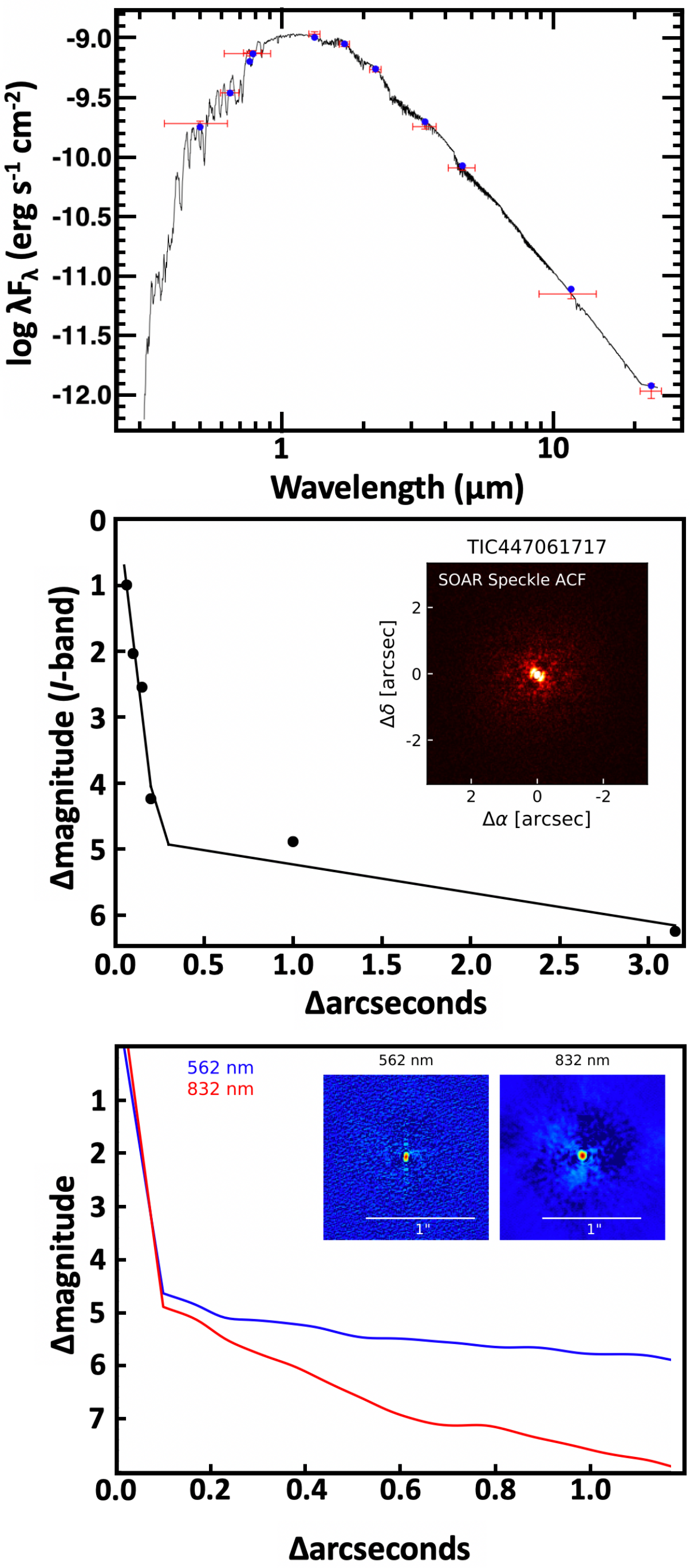}
\caption{
{\it Top}: Best fit SED for TOI-1231. The red symbols are the observed photometric measurements and the horizontal lines show
the effective width of each filter. The blue symbols depict the model flux values from the
best-fit NextGen atmosphere model, which is shown in black. 
{\it Middle}: SOAR contrast curve and image (inset) for I-band speckle observations 
of TOI-1231. 
{\it Bottom}: Zorro contrast curves and images (inset) for 562~nm and 832~nm speckle
observations of TOI-1231. No visual companions are detected in the field of
view of either SOAR or Zorro.
\label{fig:Vetting}}
\vspace{0.7cm}
\end{figure}

We analyzed TOI-1231's broadband spectral energy distribution (SED) alongside its {\it Gaia\/} EDR3 parallax to determine an empirical measurement of the star's radius, following \citet{Stassun:2016,Stassun:2017,Stassun:2018}. Together, the available photometry described here and listed in Table \ref{tab:star} cover the full stellar SED from 0.4--22~$\mu$m (see Figure~\ref{fig:Vetting}). We exclude the APASS/DR10 data from the SED fit in favor of the Gaia EDR3 passbands, which cover the same wavelength range and have smaller errors, but note that we adopt an error floor of 0.03 mag on the photometry because the systematics on the absolute flux calibration between photometric systems is 2-3\%. 

We performed the fit using NextGen stellar atmosphere models, placing a prior on the star's surface gravity ($\log g$) from the TESS Input Catalog (TIC-8).  We set the stellar effective temperature equal to the result from the Cool Dwarf Catalog \citep[\teff\, = 3557\,$\pm$\,82\,K,][]{Muirhead2018} and the extinction to zero due to the star's proximity (within the Local Bubble) which is consistent with the reddening value of $E_{B-V} = 0.001 \pm 0.015$ from \citet{Lallement2018}\footnote{STILISM: https://stilism.obspm.fr/}. The resulting SED fit is quite good (Figure~\ref{fig:Vetting}, top) with a reduced $\chi^2$ of 1.9. The best fit stellar metallicity is [Fe/H] = 0.0 $\pm$ 0.3.
Integrating the model SED results in the bolometric flux at Earth being $F_{\rm bol} = (1.345 \pm 0.047) \times 10^{-9}$ erg~s$^{-1}$~cm$^{-2}$ ($m_{bol}$ = 10.682\,$\pm$\,0.032 on IAU 2015 bolometric magnitude scale)\footnote{Besides this analysis, we also estimate
\mbol\, via other methods. Applying the M dwarf bolometric magnitude
relations of \citet{Casagrande2008} with the 2MASS $JHK_s$ photometry, 
we estimate \mbol\, = 10.74\,$\pm$\,0.02. Using the 
$V-J$ vs. $BC_{Ks}$ relation from \citet{Mann2015}
applied to the 2MASS $K_s$ magnitude we estimate \mbol\, = 
10.79. Fitting BT-Settl-CIFIST synthetic spectra to the photometry in 
Table \ref{tab:star} using the VOSA SED Analyzer \citep{Bayo2008} yielded 
$F_{\rm bol} = (1.363 \pm 0.116) \times 10^{-9}$ erg~s$^{-1}$~cm$^{-2}$ (\mbol\, = 10.666\,$\pm$\,0.095) for \teff\,=\,3600\,K, \logg\,=\,4.5, [M/H]\,=\,0. These
values provide independent checks, but we adopt the $F_{bol}$ and \mbol\, 
in the text.}. 
Taking the $F_{\rm bol}$ and $T_{\rm eff}$ together with the {\it Gaia\/} EDR3 parallax gives the stellar radius as \rstar\ = 0.466\,$\pm$\,0.021\,\rsun\ and the stellar mass as \mstar\ = 0.485\, $\pm$\, 0.024\, \msun\ based on the empirical $M_{Ks}$ vs. mass relations from \citet{Mann2019}\footnote{For $M_{Ks}$ = 5.861\,$\pm$\,0.026 and assuming [Fe/H] = 0.}. 



\subsubsection{Speckle Observations}

High-angular resolution imaging is needed to search for nearby sources that can contaminate the TESS photometry, resulting in a diluted transit and an underestimated planetary radius, and also to search for faint stars that might be responsible for the transit signal. We searched for nearby sources to TOI-1231 with SOAR speckle imaging \citep{Tokovinin2018} on UT 12 December 2019 in I-band, a similar near-IR bandpass as used by \TESS. Further details of observations from the SOAR \TESS\ survey are available in \citet{Ziegler2020}.  We detected no nearby stars within 3\arcsec\ of TOI-1231 within the 5$\sigma$ detection sensitivity limits of the observation, which are plotted along with the speckle auto-correlation function in the middle panel of Figure \ref{fig:Vetting}. Using the measured detection sensitivity from the SOAR observation and the estimated \Teff\ of the target star along with main-sequence stellar SEDs \citep{Kraus2012}, we can effectively rule out a main-sequence companion within angular resolutions between 0.2\arcsec\ to 3.0\arcsec\ or projected physical separations of 5.5 au to 83 au.

Speckle interferometric images of TOI-1231 were also obtained on UT 13 March 2020 using the Zorro\footnote{https://www.gemini.edu/sciops/instruments/alopeke-zorro/} instrument mounted on Gemini-South. Zorro observes simultaneously in two bands (832$\pm$40 nm and 562$\pm$54 nm) obtaining diffraction limited images with inner working angles of 0.026 and 0.017 arcseconds, respectively. The TOI-1231 data set consisted of 5 minutes of total integration time taken as sets of 1000 $\times$ 0.06 sec images. All the images were combined using Fourier analysis techniques, examined for stellar companions, and used to produce reconstructed speckle images \citep[see][]{Howell2011}. The speckle imaging results reveal TOI-1231 to be a single star to contrast limits of $\sim$5 to 8 magnitudes, eliminating the possibility of any main sequence companions to TOI-1231 ($d$\,=\,27.6\,pc) within the spatial limits of 3 to 33 au (Figure \ref{fig:Vetting}, bottom). 

\subsection{Stellar Kinematics and Population}

Using the astrometry and radial velocity data from \citet{GaiaDR2},
we calculate the heliocentric Galactic velocity for TOI-1231
to be ($U, V, W$) = (-20.25, -73.37, 39.45 $\pm$ 0.24, 0.36, 0.33)
\kms, with total velocity $S_{tot}$ = 85.73\,$\pm$\,0.35 \kms.
Compared to the Local Standard of Rest (LSR) of \citet{Schonrich2010},
we estimate velocities of ($U, V, W$) = (-10.2, -62.4, 46.5) \kms,
with $S_{LSR}$ = 78.4 \kms.
We use the BANYAN $\Sigma$ \citep[Bayesian Analysis for Nearby Young
AssociatioNs $\Sigma$;][]{Gagne2018} tool to estimate membership
probabilities to nearby young associations within 150 pc, however
the probabilities are $\ll$0.1\%\, for any of the known nearby stellar
groups (all with ages $<$1 Gyr), and the star is classified as ``{\it field}".
Following \citet{Bensby2014}, we use the Galactic velocity to estimate
kinematic membership probabilities to the Milky Way's principal
populations, using a 4-population model for the thin disk, thick disk, halo, and the
Hercules stream\footnote{The Hercules stream stars contain a mix of
$\alpha$-enhanced old stars and younger less $\alpha$-enhanced around
the solar [Fe/H] \citep[e.g.][]{Bensby2014}, likely from the inner part
of the Galaxy and kinematically heated by the Galactic bar
\citep{Dehnen2000} and halo, although the exact type of resonant interaction
responsible for the stream is still controversial
\citep{Monari2019}.}.
We estimate kinematic membership probabilities of $P$(thin) = 16.7\%,
$P$(thick) = 64.5\%, $P$(Hercules) = 18.7\%, and $P$(halo) = 0.06\%.
However, the star's LSR velocity places it among the Hercules stream member in
Fig.~29 of \citet{Bensby2014}.
\citet{Mackereth2018} calculated parameters of the star's Galactic
orbit using St\"{a}ckel approximation with the \citet{GaiaDR2}
astrometry, and find an eccentricity of 0.332, $z_{max}$ = 0.905 kpc,
perigalacticon of $r_{peri}$ = 4.02 kpc and apogalacticon of $r_{ap}$
= 8.03 kpc\footnote{$R_O$ = 8 kpc is assumed.}, i.e. we are catching
the star near its apogalacticon.

We also searched for companions of TOI-1231 in the \citet{GaiaDR2} catalog via the 50\,pc sample of \citet{Torres19}. 
Given the mass of 0.46\,\msun, we estimate the tidal radius for TOI-1231 (where bound companions would likely be found) to be 1.04\,pc \citep{Mamajek13}, which corresponds to a projected radius of $\sim$2$^{\circ}$.2.
Querying the \citet{Torres19} catalog for stars with proper motions and parallaxes within 20\%\, of that of the star within 2 tidal radii (4$^{\circ}$.4) yielded no candidate companion.
Therefore, we conclude TOI-1231 to be a single star.

\subsection{Metallicity}\label{sec:metallicity}

The $V-K_s$ vs. absolute magnitude position of an M dwarf can be used 
to infer a photometric metallicity estimate \citep[e.g.][]{Johnson2009}.
TOI-1231's combination of $V-K_s$ color (4.25) and absolute magnitude
($M_{Ks}$ = 5.86) are consistent with it being 0.17 mag brighter
than the locus for nearby M dwarfs, which \citet{Schlaufman2010} estimate
represents an isometallicity trend of [Fe/H] = -0.14. 
Using the calibrations of \citet{Johnson2009} and \citet{Schlaufman2010},
this offset is consistent with a predicted metallicity of 
[Fe/H] = +0.05 and -0.03, respectively, i.e., approximately solar. 

Gaia DR2 \citep{GaiaDR2} has an estimate of [Fe/H]\,=\,-1.5, but for an unrealistic giant-like surface gravity of \logg\,=\,3.0 and hot \teff\, of 4000\,K. 
Taken at face value, the Gaia DR2 metallicity would predict that
the star's $V-K_s$ vs. $M_{Ks}$ position should be more than a magnitude
below the main sequence \citep[extrapolating the metallicity vs. $\Delta M_{Ks}$ relations of][]{Schlaufman2010}, well below where it is observed ($\sim$0.2 mag above the MS).

\citet{Anders19} uses the \emph{StarHorse} code to fit $G_B,\ G,\ G_R,\ J,\ H,\ Ks,\ W1,\ \rm{and}\ W2$ photometry to solve for a photometric metallicity estimate consistent with [Fe/H] = 0.095$^{+0.028}_{-0.061}$. 

Taking the mean of our independent photometric metallicity estimates and that of \citet{Anders19}, we adopt a metallicity of [Fe/H] = +0.05\,$\pm$\,0.08. 

\section{Exoplanet Detection \& Follow Up \label{sec:planetdata}}

\subsection{\TESS\ Time Series Photometry}

\thisstar\ was selected for transit detection observations by \TESS\ from two input lists.  It was included in the exoplanet candidate target list (CTL) that accompanied version 8 of the TESS Input Catalog \citep[TIC; ][]{Stassun2019}, and also in the Cool Dwarf Catalog \citep{Muirhead2018}.  Its CTL observing priority was 0.00734, placing it among the top 3\% of targets selected for transit detection by the mission, due to its brightness and small estimated stellar radius (see sections 3.1 and 3.3 of \citet{Stassun2019} for details on the prioritization process). It was also selected for observations by TESS Guest Investigator proposal GO11180 (C. Dressing). \thisstar\ was observed by \TESS\ from UT 28 February through 26 March 2019 as part of the Sector 9 campaign and again from UT 26 March through 22 April 2019 as part of Sector 10. The star fell on Camera~3 in both sectors, but shifted from CCD~1 in Sector~9 to CCD~2 in Sector~10.

\begin{figure*}
\includegraphics[width=\linewidth]{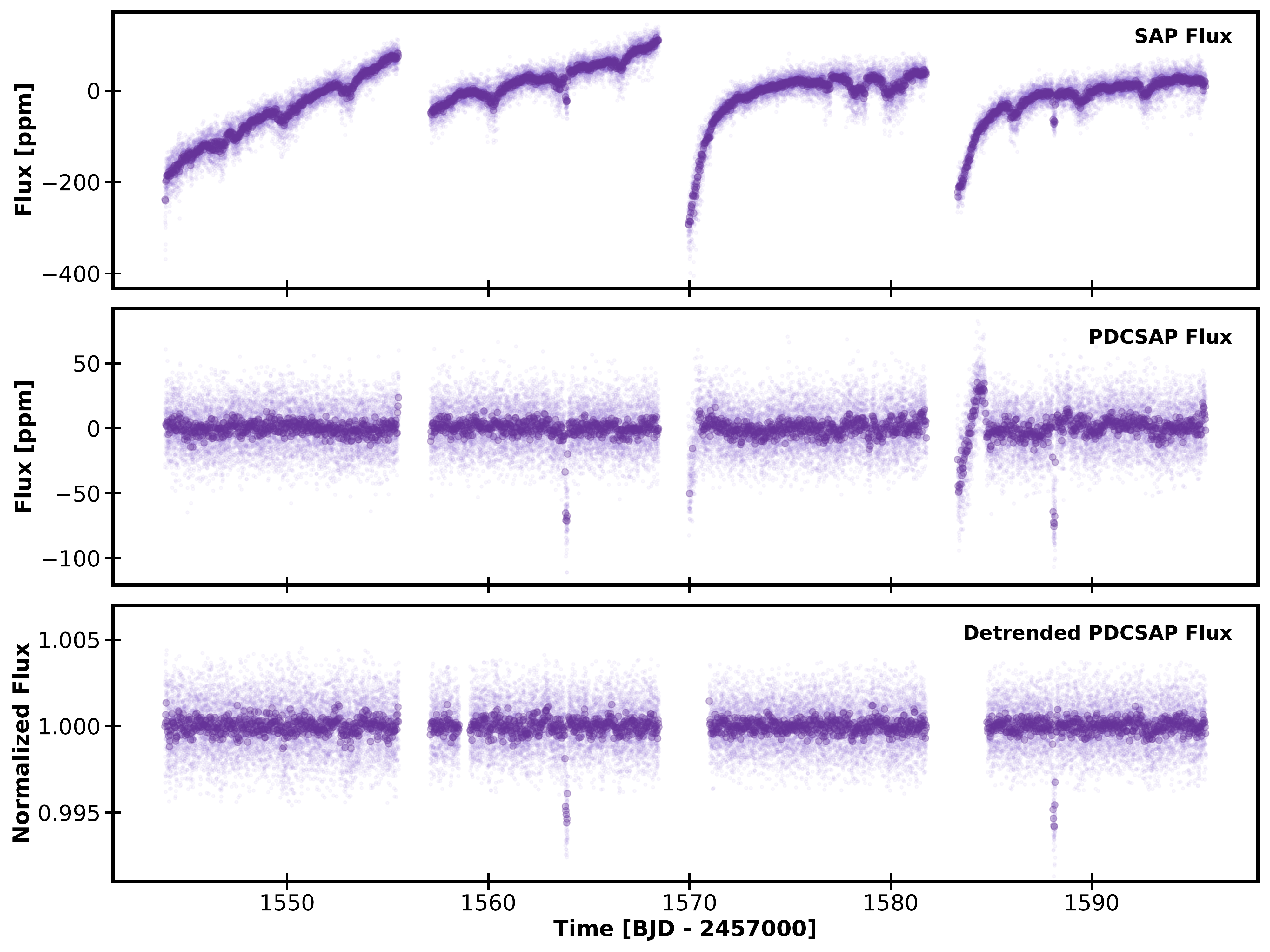}
\caption{TESS light curves: Simple Aperture Photometry (\emph{SAP, top}), presearch data conditioned SAP (\emph{PDCSAP, middle}), and orbit-by-orbit detrended PDCSAP (\emph{bottom}). Lighter points depict the \TESS\ two-minute cadence flux measurements, darker points are the same data binned into 30 minute intervals. The two transit events can be clearly seen as brief flux dips in the middle and bottom panels.
\label{fig:lightcurve}}
\vspace{0.7cm}
\end{figure*}

The SPOC data for \thisstar\, can be accessed at the the Mikulski Archive for Space Telescopes (MAST) website\footnote{https://mast.stsci.edu}, and includes both the simple aperture photometry (SAP) flux measurements \citep{Twicken2010, Morris2017} and the presearch data conditioned simple aperture photometry (PDCSAP) flux measurements \citep{Smith2012, Stumpe2012, Stumpe2014}. These data products differ in that the instrumental variations present in the SAP measurements are removed from the PDCSAP data. The main variations are due to thermal effects and strong scattered light present at the start of each orbit, which impact the systematic error removal in PDC (see \TESS\ data release notes\footnote{https://archive.stsci.edu/tess/tess\_drn.html} DRN16 and DRN17). We therefore use the quality flags provided by SPOC to mask out unreliable segments of the time series before executing the global fitting process in Section \ref{sec:exofastv2}. We further detrend the \TESS\ data by separating the individual spacecraft orbits (two in each sector) and fitting each orbit's flux measurements with a low order spline in order to mitigate residual trends in the photometry (Figure \ref{fig:lightcurve}).

\subsubsection{TESS Transit Detection}

\thisplanetb\ transited once in each of sectors 9 and 10. It was first identified as a planet candidate five months prior to becoming a TOI, by the TESS Single Transit Planet Candidate Working Group (TSTPC WG). The TSTPC WG focuses on searching light curves produced by the MIT Quick Look Pipeline for single transit events, and validating and/or confirming those that are true planets, with the aim of increasing the yield of intermediate-to-long-period planets found by \TESS\ (\citealt{Villanueva:2019}, Villanueva et al. in prep.). 

The two transits of \thisplanetb\ were later also detected by both the MIT Quick Look Pipeline (QLP), which searches for evidence of planet candidates in the \TESS\ 30 minute cadence Full Frame Images, and the SPOC pipeline, which analyzes the 2-minute cadence data that \TESS\ obtains for pre-selected target stars \citep{Jenkins:2016}. The \TESS\ transits, one of which occurs in Sector 9 and the other in Sector 10, have a measured depth of 6453 ppm, a duration of 3.26 hours, and a measured period of 24.246 days. 

While the depth and flat bottomed shape of the TOI-1231 transits were suggestive of the transit signal being planetary in nature, there are a variety of false positives that can mimic this combination. The main source of false positives in the TESS Objects of Interest (TOIs) are eclipsing binaries, either as two transiting stars on grazing orbits or in the case of a background blend which reduces the amplitude of a foreground eclipsing binary signal, causing it to be fallaciously small \citep[e.g.][]{Cameron2012}. The \TESS\ vetting process is designed to guard against these false positives, and so we inspected the star's Data Validation Report \citep[DVR,][]{Twicken2018,Li2019}, which is based upon the SPOC two minute cadence data. The multi-sector DVR shows no signs of secondary eclipses, odd/even transit depth inconsistencies, nor correlations between the depth of the transit and the size of the aperture used to extract the light curve, any of which would indicate that the transit signal originated from by a nearby eclipsing binary. The DVR also showed that the location of the transit source is consistent with the position of the target star. Upon passing these vetting checks, the transit signal was assigned the identifier TOI-1231.01 and announced on the MIT \TESS\ data alerts website\footnote{http://tess.mit.edu/alerts} \citep{Guerrero2021}.

\subsection{Ground-based Time-Series Photometry}

We acquired ground-based time-series follow-up photometry of \thisstar\ during the times of transit predicted by the \TESS\ data. We used the {\tt TESS Transit Finder}, which is a customized version of the {\tt Tapir} software package \citep{Jensen:2013}, to schedule our transit observations.

\subsubsection{LCO 1m Observations}
Two partial transits of \thisplanetb\ were observed using the Las Cumbres Observatory Global Telescope (LCOGT) 1m network \citep{Brown:2013} in the Pan-STARSS $z$ band on UTC 2020 January 16 by the LCOGT node at Cerro Tololo Inter-American Observatory and May 6 2020 by the LCOGT node at Siding Spring Observatory (Figure \ref{fig:FitResults}, second panel). The telescopes are equipped with $4096\times4096$ LCO SINISTRO cameras having an image scale of 0$\farcs$389 pixel$^{-1}$ resulting in a $26\arcmin \times 26\arcmin$ field of view. The images were calibrated by the standard LCOGT BANZAI pipeline and the photometric data were extracted using the {\tt AstroImageJ} ({\tt AIJ}) software package \citep{Collins:2017}. Circular apertures with radius 12 pixels (4$\farcs$7) were used to extract the differential photometry.

\subsubsection{ASTEP 0.4m Observations}
We observed two full transits of \thisstar\ with the Antarctica Search for Transiting ExoPlanets (ASTEP) program on the East Antarctic plateau \citep{Guillot2015,Mekarnia2016}. The $0.4$m telescope is equipped with an FLI Proline science camera with a KAF-16801E, $4096\times4096$ front-illuminated CCD. The camera has an image scale of 0$\farcs$93 pixel$^{-1}$ resulting in a $1^{\circ}\times1^{\circ}$ corrected field of view. The focal instrument dichroic plate splits the beam into a blue wavelength channel for guiding, and a non-filtered red science channel roughly matching an R$_{c}$ transmission curve. The telescope is automated or remotely operated when needed. Due to the extremely low data transmission rate at the Concordia Station, the data are processed on-site using an automated IDL-based pipeline, and the result is reported via email and then transferred to Europe on a server in Roma, Italy. The raw light curves of about 1,000 stars are then available for deeper analysis. These data files contain each star's flux computed through various fixed circular apertures radii, so that optimal lightcurves can be extracted (Figure \ref{fig:FitResults}, third panel). For \thisstar\ an 11 pixels (10$\farcs$3) radius aperture was found to give the best results.

The observations took place on UTC 2020 May 6 and August 11. Weather was good to acceptable, and air temperatures ranged between $-50^\circ$C and $-70^\circ$C.  Two full transits, including the ingress and egress, were detected. Two other transits of TOI-1231~b were also detected on May 30 and June 26, but were partial or affected by technical issues, are generally of lower signal-to-noise and are thus not included in the present analysis. In each case, the ingress and egress occurred at the predicted times (with an uncertainty of a few minutes or less), indicating that any transit time variation must be small. 

\subsubsection{ExTrA 0.6m Observations}\label{subsection:extra}
\begin{figure}
\includegraphics[width=.48 \textwidth]{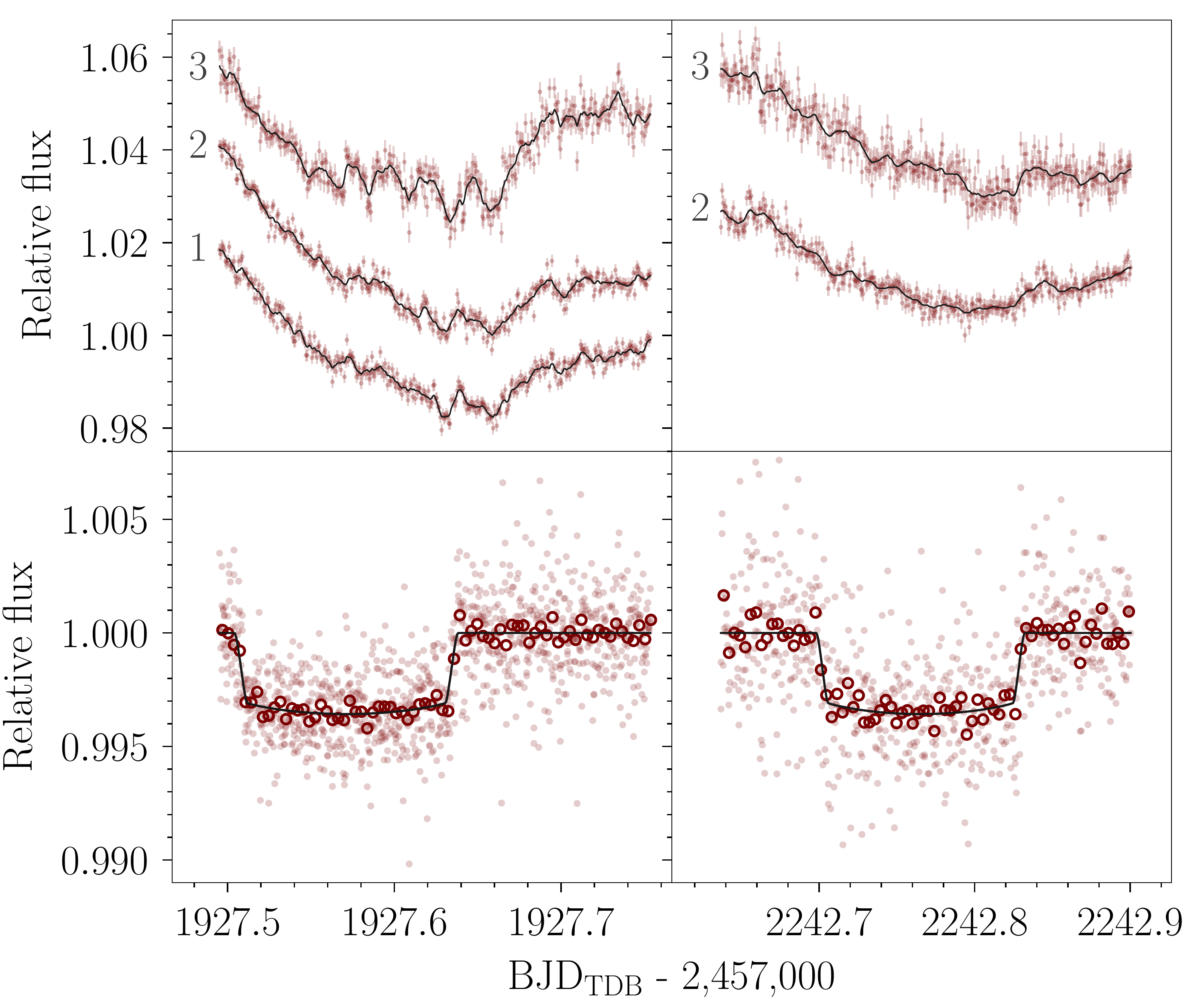}
\caption{Light curves obtained by the ExTrA facility on nights UTC 2020 March 18 (left) and 2021 January 27 (right). Top panels show the Raw ExTrA photometry of each telescope (labeled 1, 2, and 3), displaced vertically for clarity and a black curve depicting the median of the posterior of the modeling. Bottom panels show the light curves after correction with the maximum a posteriori model, with individual observations as light points, 5 minute bins as dark circles, and the maximum a posteriori transit model as a black line.
\label{fig:ExTrA}}
\end{figure}

The ExTrA facility \citep{bonfils2015}, located at La Silla observatory, consists of a near-infrared (0.85 to 1.55~$\mu$m) multi-object spectrograph fed by three 60-cm telescopes. At the focal plane of each telescope five fiber positioners pick the light from the target and four comparison stars. We observed two full transits of \thisplanetb\ on UTC 2020 March 18 and 2021 January 27 (Figure~\ref{fig:ExTrA}). The first night we observed with three telescopes using the fibers with 8\arcsec\ apertures. The second night we observed with two telescopes and 4\arcsec\ aperture fibers. Both nights we used the high resolution mode of the spectrograph ($R\sim$200) and 60-seconds exposures. We also observed 2MASS J10283882-5234151, 2MASS J10285562-5220140, 2MASS J10273532-5206553, and 2MASS J10271569-5239330, with J-magnitude \citep{2mass} and T$_{\rm eff}$ \citep{GaiaDR2} similar to \thisstar, for use as  comparison stars. The resulting ExTrA data were analysed using custom data reduction software.

The ExTrA light curves are affected by systematic effects that are currently under investigation. To account for them, we modeled the transits observed by ExTrA with {\sc \tt juliet} \citep{espinoza2019,kreidberg2015,speagle2020}, and included the quasi-periodic kernel Gaussian Process implemented in {\sc \tt celerite} \citep{foreman-mackey2017}, with different kernel hyperparameters for each ExTrA telescope and for each night. We used a prior for the stellar density of $\rhostar=6.48\pm0.30~\gcmc$ \citep{Stassun2019}, and non-informative priors for the rest of the parameters. The posterior provides the timings of the observed transits: $2458927.5716\pm0.0013$~BJD$_{\rm TDB}$, and $2459242.7657\pm0.0013$~BJD$_{\rm TDB}$, a planet to star radius ratio of $R_P/R_*=0.0660_{-0.0086}^{+0.0072}$, and an impact parameter of $b=0.201_{-0.12}^{+0.091}$. These transit light curves are not used in the global EXOFATv2 fit described in Section~\ref{sec:exofastv2} in order to avoid biasing the final results due to the systematics present in the ExTrA data. However, the timing of the second transit is used as a prior for the time of conjunction ($T_C$) to better constrain the period of the planet.

\subsection{Time Series Radial Velocities}

Shortly after the discovery of the planet candidate by the TSTPC WG, we began radial velocity (RV) follow up efforts using the Planet Finder Spectrograph (PFS) on Las Campanas Observatory's 6.5m Magellan Clay telescope \citep{Crane2006, Crane2008, Crane2010}. PFS is an iodine cell-based precision RV spectrograph with an average resolution of $R \simeq$ 130,000. RV values are measured by placing a cell of gaseous I$_2$, which has been scanned at a resolution of 1 million using the NIST FTS spectrometer \citep{Nave2017}, in the converging beam of the telescope. This cell imprints the 5000-6200\AA\ region of the incoming stellar spectra with a dense forest of I$_2$ lines that act as a wavelength calibrator and provide a proxy for the point spread function (PSF) of the spectrometer \citep{MarcyButler1992}. 

The spectra are split into 2\AA\ chunks, each of which is analyzed using the spectral synthesis technique described in \citet{Butler1996}, which deconvolves the stellar spectrum from the I$_2$ absorption lines and produces an independent measure of the wavelength, instrument PSF, and Doppler shift. The  final Doppler velocity from a given observation is the weighted mean of the velocities of all the individual chunks ($\sim$800 for PFS). The final internal uncertainty of each velocity is the standard deviation of all 800 chunk velocities about that mean. 

\begin{figure}
\includegraphics[width=.45 \textwidth]{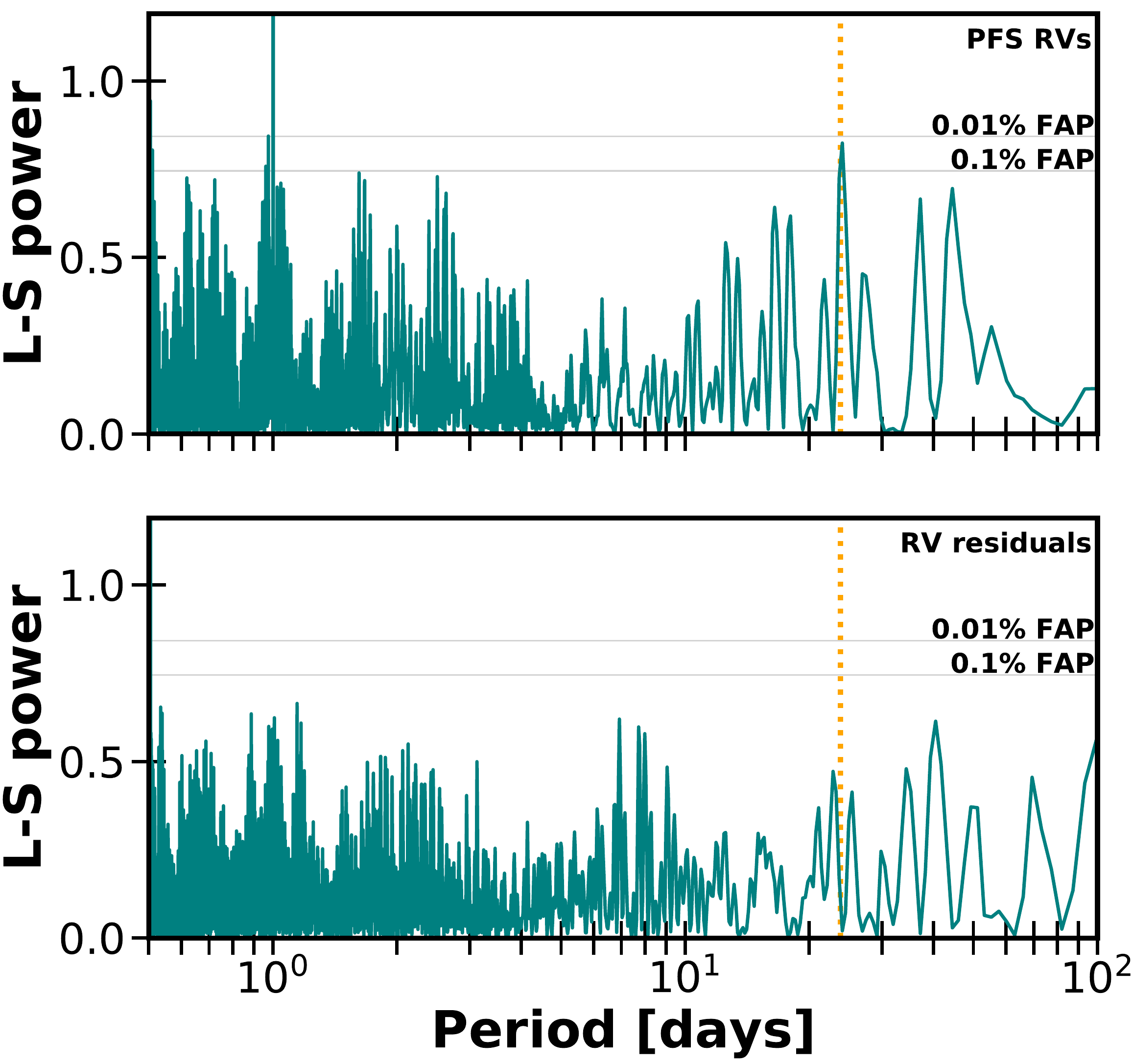}
\caption{
{\it Top}: Generalized Lomb-Scargle periodogram of the radial velocity measurements from PFS. There is a clear peak at period of TOI-1231~b (orange dotted line).
{\it Bottom}: GLS of the radial velocity residuals, after the signal for planet b has been removed. No significant peaks remain in the data to suggest the presence of additional planets and/or stellar rotational signals.
\label{fig:RVPeriodograms}}
\vspace{0.7cm}
\end{figure}

A total of 28 PFS radial observations were obtained from May 2019 to February 2020, binned into 14 velocity measurements, with a mean internal uncertainty of 1.22~\ms (Table \ref{tab:RVs}). A Generalized Lomb-Scargle (GLS) periodogram of the PFS RV data shows a significant peak at 24 days (Figure \ref{fig:RVPeriodograms}), which matches the orbital period determined from the \TESS\ data.

\begin{table}[htbp]
\centering
\caption{Binned RV data of TOI-1231}
\begin{tabular}{ccc}
\hline\hline
Date [BJD$_{\rm{TDB}}$] & RV [$\ms$]& $\sigma_{RV}$ [$\ms$] \\
\hline \vspace{2pt}
2458618.501473 & -10.13 & 1.19 \\
2458625.551234 & 0.32 & 1.44 \\
2458627.574495 & 3.73 & 1.17 \\
2458677.491563 & 7.16 & 1.27 \\
2458679.483330 & 5.96 & 1.22 \\
2458685.500395 & 2.12 & 1.44 \\
2458827.826089 & 3.29 & 1.06 \\
2458828.842260 & -0.34 & 1.08 \\
2458833.804469 & -9.11 & 1.27 \\
2458836.814827 & -4.43 & 1.01 \\
2458883.799738 & -4.85 & 1.21 \\
2458885.807135 & -0.57 & 1.24 \\
2458886.798816 & -2.90 & 1.31 \\
2458890.781648 & 1.10 & 1.17 \\
\hline\hline
\end{tabular}
\label{tab:RVs}
\end{table}


\section{System Parameters from EXOFASTv2}
\label{sec:exofastv2}

To fully characterize the \thisstar\ system, we used the EXOFASTv2 software package \citep{Eastman2013, Eastman2019} to perform a simultaneous fit to the star's broadband photometry, the \TESS, LCO, and ASTEP time series photometry, and the PFS radial velocity measurements. We applied Gaussian priors to the parallax and V-band extinction of the star using the results of Gaia EDR3 ($\varpi = 36.3726\pm0.0163$, corrected using the \citet{Lindegren2020} prescription) and \citet{Lallement2018} (A$_v$=$0.003\pm0.0465$), respectively. Gaussion priors were also placed on the TOI-1231's mass (\mstar = 0.461 $\pm$ 0.018 \msun, calculated using the prescription in \citealt{Mann2019}), effective temperature (T$_{\rm{eff}}$=3562$\pm$101 K, taken from \citealt{Gaidos2014}) and metallicity ([Fe/H] = 0.05 $\pm$ 0.08 (see S\ref{sec:metallicity} for details). We disabled EXOFASTv2's MIST isochrone fitting option, which is less reliable for low mass stars \citep{Eastman2019}. Finally, we placed a prior on the planet's time of conjunction derived from the ExTrA photometry (\ref{subsection:extra}). We did not place any constraints on the planet's period or orbital eccentricity.

EXOFASTv2's SED fitting methodology differs from the approach used in
the SED-only fit that helped verify TOI-1231's suitability for PRV
follow up in Section \ref{sec:SED}. In place of the NextGen
atmospheric models, EXOFASTv2 instead uses pre-computed bolometric
corrections in a grid of \logg , \Teff , [Fe/H], and V-band extinction
\footnote{\url{http://waps.cfa.harvard.edu/MIST/model_grids.html##bolometric}}. This grid is based on the ATLAS/SYNTHE stellar atmospheres \citep{Kurucz2005} and the detailed shapes of the broadband photometric filters.

We note that neither the raw TESS time series photometry nor the PFS RV measurements exhibit the type of sinusoidal variations that we would expect to see if the star was subject to rotation-based activity due to active regions such as star spots or plages crossing the visible hemisphere \citep{Saar1997, Robertson2020}. This lack of rotational modulation suggests an inactive star, a claim further supported by the lack of emission or any detectable temporal changes in the core of the H-$\alpha$ line \citep[see, e.g., ][]{Reiners2012, Robertson2013}. We therefore do not include any additional activity-based terms or detrending efforts when fitting either the photometric or radial velocity data.

\begin{figure}[ht]
\includegraphics[width=0.45\textwidth]{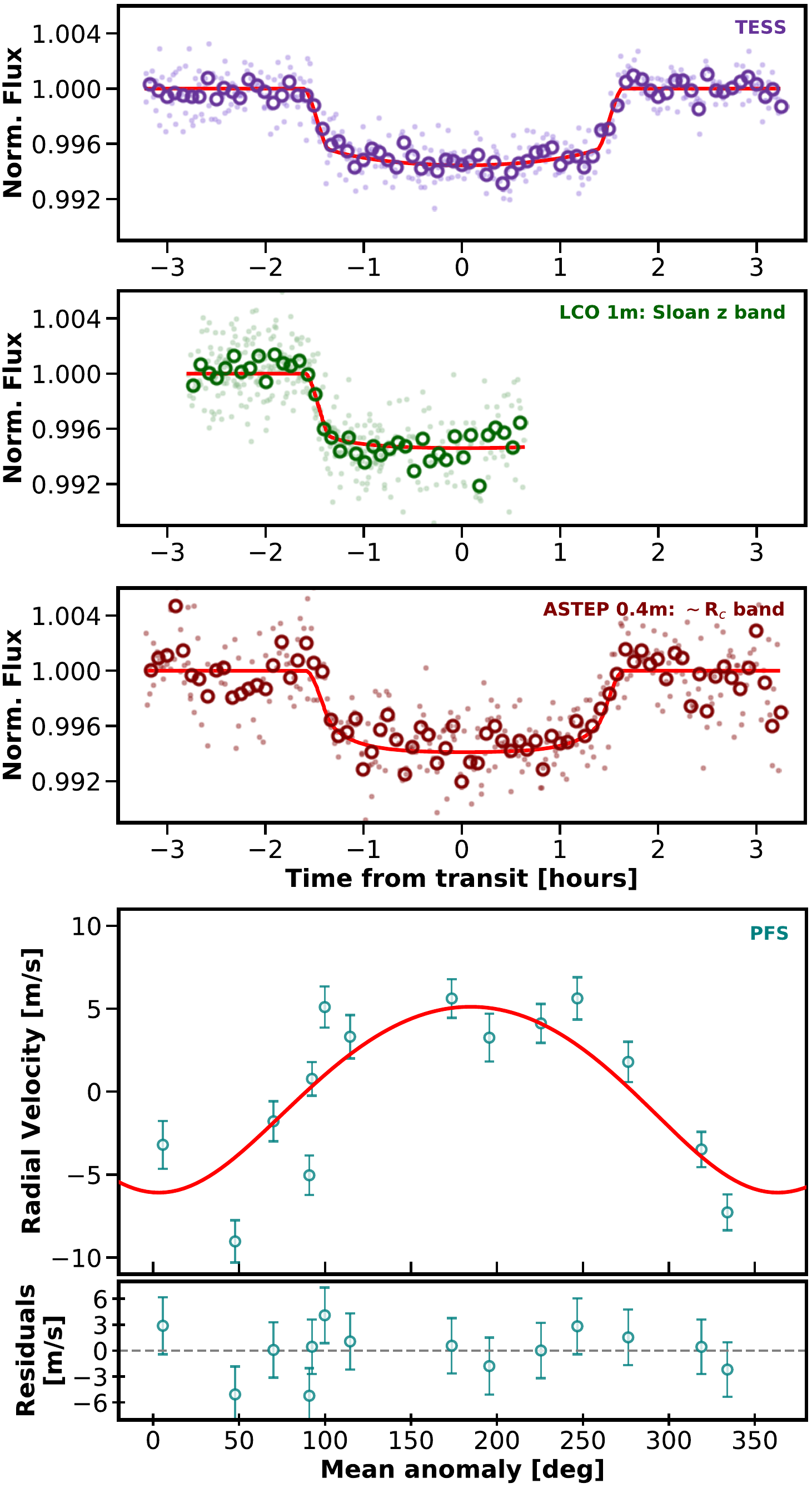}
\caption{Results of the EXOFASTv2 joint fit to the TESS photometry, ground-based LCO and ASTEP, and the PFS radial velocities. The top three panels show the phase folded photometry data from TESS (two full transits), LCO (two partial transits), and ASTEP (two full transits). All three photometry panels depict the unbinned data as light points, the five minute binned data as dark circles, and the best fit model as a solid red line. The bottom panel shows the phase folded RV measurements from PFS in cyan and the best fit model in red. The RV residuals after the best fit model has been subtracted from the data are displayed underneath the fit. Error bars in both RV panels are the quadrature sum of the PFS internal uncertainties and the RV jitter estimate from the EXOFASTv2 fit.
\label{fig:FitResults}}
\vspace{0.7cm}
\end{figure}

The median EXOFASTv2 parameters for the \thisstar\ system are shown in Table 3 and the best fits to the TESS, LCO, and ASTEP photometry and the PFS radial velocity data are shown in Figure \ref{fig:FitResults}. The scatter in a star's radial velocity measurements includes any unmodeled instrumental effects or stellar variability. To address this, we include a `jitter' term in the RV fit which is used to encompass uncorrelated signals in the star's own variability or PFS's systematics that occur on timescales shorter than the observational baseline. This value is added in quadrature to the internal uncertainties reported in the PFS data set to produce the RV error bars seen in Figure \ref{fig:FitResults}.  The best fit orbital eccentricity ($e$ = 0.087$^{+0.012}_{-0.061}$)  should not be regarded as statistically significant as it does not meet the criteria of being at least 2.45$\sigma$ from 0 that is necessary to avoid falling subject to the Lucy-Sweeney bias \citep{LucySweeney1971}. Even though the best fit eccentricity is consistent with a circular orbit, we do not enforce a zero eccentricity fit because even a small amount of non-modeled eccentricity can bias the resulting orbital parameters and underestimate the uncertainties in many covariant parameters.

The mass of \thisplanetb\ is measured to be \plmassunc, which, when combined with the measured planet radius of \plradunc, results in a bulk density of \plrhounc making the planet slightly denser than Neptune ($\rho_{Nep} = 1.638$ g cm$^{-3}$).


\section{Discussion}
\label{sec:discussion}
\begin{figure}[ht]
\includegraphics[width=\linewidth]{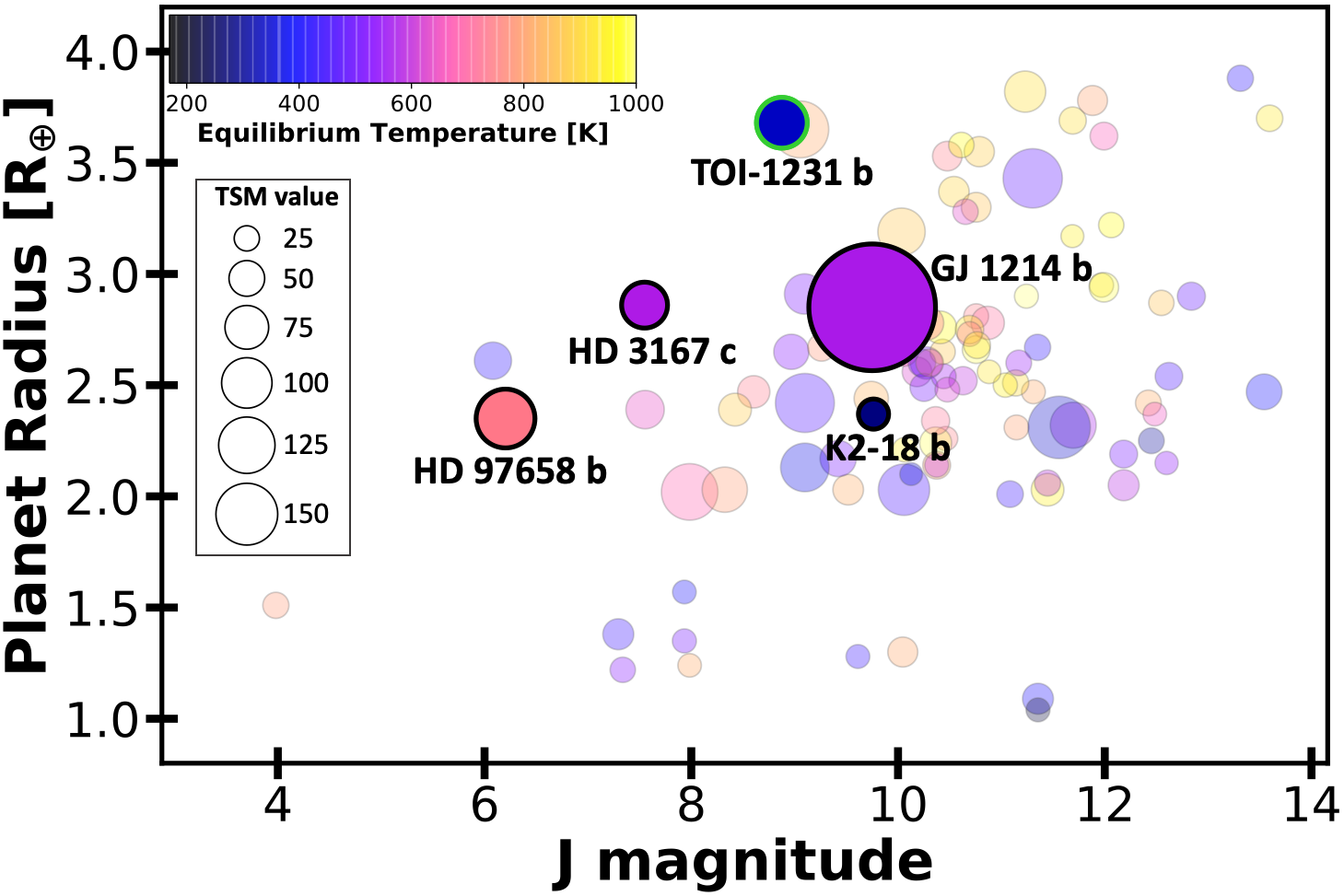}
\caption{Transmission spectroscopy metric (TSM) values for small planets with T$_{\rm{eq}}$ \textless 1000K compared to their radii and host star's J magnitude, data are taken from \citet{Guo2020}. The four planets that have already had their atmospheres characterized by HST are shown as solid circles and labeled. TOI-1231 b (green outline) offers another promising target in the cool, small planet regime and would, for the first time, enable comparisons in the T$_{\rm{eq}}$ = 250-350 K range when contrasted with K2-18 b.
\label{fig:TSM_fig}}
\end{figure}

\cite{Barclay18} predict that \TESS\, will find just one non-rocky Neptune-sized or smaller planet in the habitable zone of a M dwarf brighter than $J$ = 10. While the final number may be slightly higher, TOI-1231~b is the first confirmed \TESS\ planet to meet these criteria. One method for comparing planets' potential for atmospheric characterization via transmission spectroscopy is the Transmission Spectroscopy Metric \citep[TSM, ][]{Kempton:2018}. A planet's TSM is proportional to its transmission spectroscopy signal-to-noise and is based on the strength of its expected spectral features (derived from its radius and scale height and the radius of the host star) and the host star's apparent J-band magnitude. With a transmission spectroscopy metric (TSM) of 99 $\pm$ 25, TOI-1231~b planet ranks among the highest TSM Neptunes of any temperature in the pre-TESS era \citep[Figure \ref{fig:TSM_fig}, ][]{Guo2020}.

TOI-1231~b is one of the coolest planets accessible for atmospheric studies with T$_{eq}$ = 330 K\footnote{We note that T$_{eq}$ is calculated using Eqn. 1 of \citet{HansenBarman2007}, which assumes no albedo and perfect redistribution. As such the quoted statistical error is likely severely underestimated relative to the systemic errors inherent in this assumption.}. Until recently it appeared that cooler planets had smaller spectral features, perhaps due to the increasing number of condensates that can form at lower temperatures \citep{CrossfieldKreidberg2017}. However, new observations of water features in the habitable-zone planet K2-18~b break this trend \citep{tsiaras19, benneke19}. The K2-18~b water feature is very intriguing: it is suggestive of a qualitative change in atmospheric properties near the habitable zone. Perhaps condensates rain out (analogous to the L/T transition in brown dwarfs), and/or photochemical haze production is less efficient \citep{saumon08, morley13}. However, K2-18~b is the only planet below 350 K with a measured transmission spectrum. TOI-1231~b provides an intriguing addition to the atmospheric characterization sample in this temperature range to determine whether K2-18~b is representative or an outlier. Recently, four HST transit observations were awarded to measure the near-infrared transmission spectrum of TOI-1231 b with the Wide Field Camera 3 (WFC3) instrument (GO 16181; PI L. Kreidberg).

\subsection{Simulated Atmospheric Retrievals}

In order to estimate how well the atmospheric properties could be extracted with HST, we used the open-source \texttt{petitRADTRANS} package \citep{mollierewardenier2019} to derive transmission spectra of TOI-1231~b, based on a simple atmospheric model. The atmosphere probed by the observations was assumed to be isothermal, at the equilibrium temperature derived for the planet in this work. Next, equilibrium chemistry was used to calculate the absorber abundances in the atmosphere, obtained with the chemistry model that is part of \texttt{petitCODE} \citep{mollierevanboekel2017}. We assumed two different compositional setups, 3 and 100~$\times$~solar (Jupiter and Neptune-like, respectively), at a solar C/O. In addition, we introduced a gray cloud deck and modeled its effect on the spectrum when placing it between 100 and $10^{-6}$~bar, in 1~dex steps. The model with the highest cloud pressure was assumed to be our cloud-free model, because the atmosphere will become optically thick at lower pressures. We included the gas opacities of the following line absorbers: H$_2$O, CH$_4$, CO, CO$_2$, Na, and K. In addition to the gray cloud, continuum opacity sources arising from H$_2$, He, CO, H$_2$O, CH$_4$, and CO$_2$ Rayleigh scattering, as well as H$_2$-H$_2$ and H$_2$-He collision-induced absorption, were included. We refer the reader to \citet{mollierewardenier2019} for the references used for the opacities.

\begin{figure}
\includegraphics[width=\linewidth]{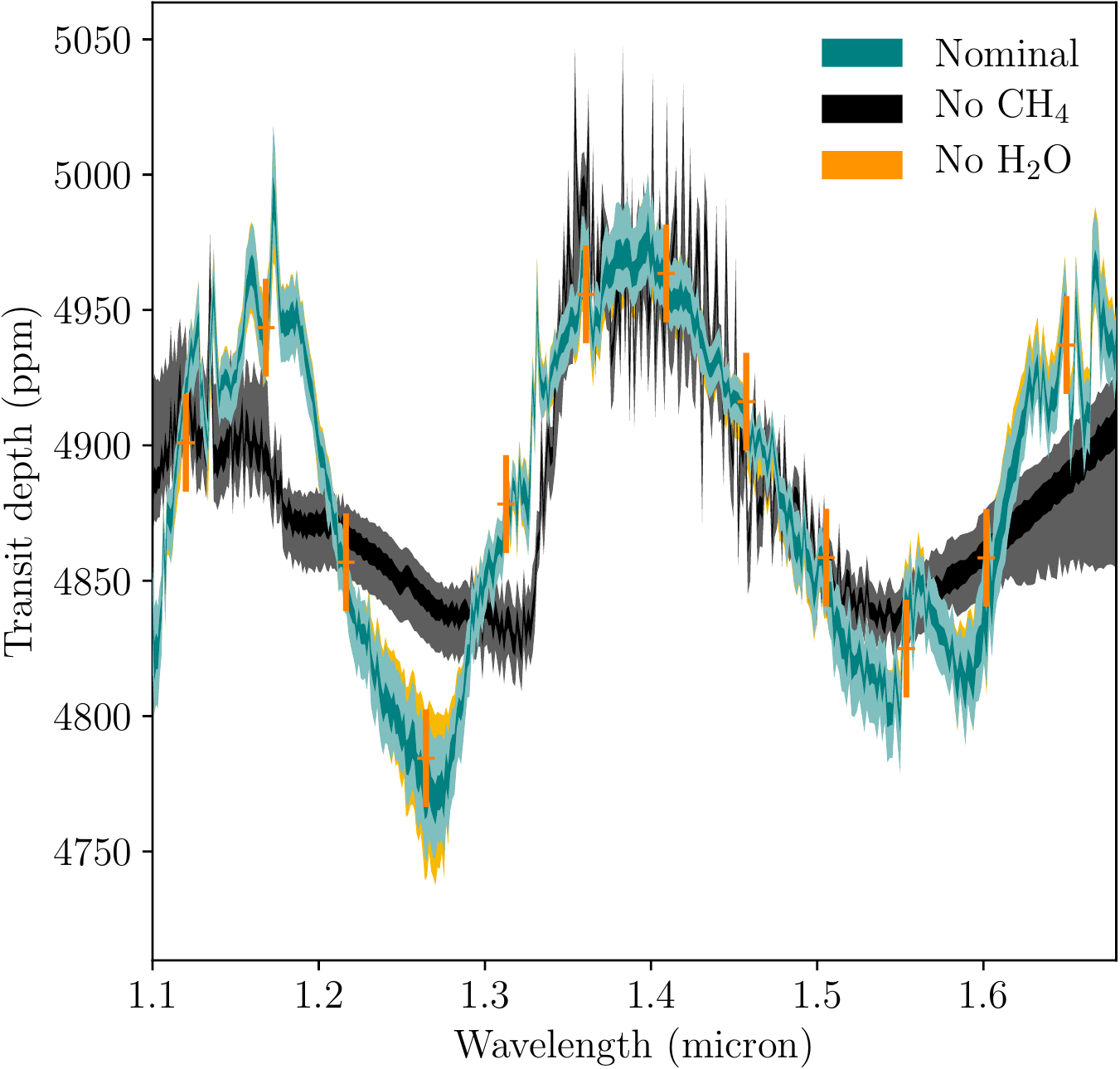}
\caption{Synthetic HST observations and retrieval of the clear, 100~$\times$~solar enrichment case. The 16-18 and 2-98 percentile envelopes of the retrieved flux distribution are shown for the retrievals with the full model (green envelopes) and the model neglecting CH$_4$ or H$_2$O (black or orange envelopes, respectively).
\label{fig:RetrievalFig}}
\end{figure}

We generated mock observations for all cases described above and retrieved them with \texttt{petitRADTRANS}, using the \texttt{PyMultiNest} package \citep{buchnergeorgakakis2014}. The latter uses the nested sampling implementation \texttt{MultiNest} \citep{ferozhobson2009}. The synthetic HST WCF3 observations were created assuming a wavelength range of 1.12 to 1.65~$\mu$m, with 12 points spaced equidistantly in wavelength space. We estimated the uncertainties on the spectroscopic transit depths using the \texttt{Pandexo\_HST} tool\footnote{\url{https://exoctk.stsci.edu/}}. Assuming four HST transit observations, we expect uncertainties of 18 ppm on the transit in each spectral channel. This corresponds to $\sim 0.6$ times the transit signal of the planet's scale height, when assuming a 100~$\times$~solar composition. For these retrievals we placed special emphasis on the detectability of H$_2$O and CH$_4$, for which we implemented the method described in \citet{bennekeseager2013}: the abundances of all metal absorbers were retrieved freely, assuming vertically constant abundance profiles. The abundance of H$_2$ and He was found by requiring that the mass fractions of all species (metals + H$_2$ and He) add up to unity, with an abundance ratio of 3:1 between H$_2$ and He. Three retrievals were run for every synthetic observation. (i): nominal model, retrieving the abundances of all metal absorbers, as well as the cloud deck pressure. (ii): same as (i), but neglecting the CH$_4$ opacity and CH$_4$ abundance as a free parameter. (iii): same as (i), but neglecting the H$_2$O opacity and H$_2$O abundance as a free parameter.

For every synthetic observation we then derived three evidences for models (i), (ii), and (iii), using nested sampling retrievals. The Bayes factor $B$, which is the ratio of these evidences, then allows us to assess how strongly models including CH$_4$ or H$_2$O are preferred (Bayes factor of models (i) and (ii) or models (i) and (iii), respectively). We used a boundary value of $B>3$ to express strong preference for a given model, following \citet{kass95}.

We show the synthetic HST observation of the clear, 100~$\times$~solar case in Figure~\ref{fig:RetrievalFig}. Because the 100~$\times$~solar case has a larger mean molecular weight, it is the more challenging of the two enrichment cases. In addition to the synthetic observation, the 16-18 and 2-98 percentile envelopes of the retrieved transit depth distribution are shown for the retrievals with models (i), (ii), and (iii), that is the full model and the model neglecting the CH$_4$ or H$_2$O opacity, respectively. 

For these cases, we find very strong preference for including CH$_4$ in the model ($B=10^8$) and no preference for including H$_2$O (H$_2$O and CH$_4$ have roughly equal abundances in the input model but the CH$4$ opacity is larger than the H$_2$O opacity at all wavelengths).  Clouds are a possibility for such planets \citep{CrossfieldKreidberg2017} and we find that from $P_{\rm cloud} < 1$~mbar on it becomes challenging to detect atmospheric features at all, and the three models (i), (ii) and (iii) become indistinguishable. In summary, we conclude that TOI-1231~b is an excellent target for atmospheric characterization. With a few transit observations, it will be possible to detect spectral features in an atmosphere similar to that of K2-18~b \citep{benneke19,tsiaras19}, enabling the first comparative planetology in the temperature range $250 - 350$ Kelvin.

\subsection{Prospects for \jwst\ Observations}

With just one transit observed with \jwst's NIRISS, for a clear, solar composition atmosphere, we expect to detect \thisplanetb's spectrum (dominated by water at NIRISS wavelengths) with 90$\sigma$ significance \citep{Kempton:2018}. 

We also investigate prospects for a cloudy atmosphere, using PLATON \citep{Zhang2018} to generate a solar composition model spectrum with a cloud deck pressure of 10 mbar (similar to GJ~3470\,b; \citealt{Benneke2019}). We then used PandEXO \citep{Batalha2017} to simulate a transmission spectrum for such an atmosphere. We find that even in this scenario, one transit with the NIRISS instrument would be sufficient to detect water absorption with 7.5$\sigma$ significance. For reference, this is 2$\sigma$ higher than the detection significance obtained with {\it six} HST WFC3 transits for GJ 3470b \citep{Benneke2019}, a planet with a similar size but much lower density (and also orbiting a 0.5 \rsun\ star).

\subsection{Probing Atmospheric Escape}

Given TOI-1231~b's low gravitational potential and expected XUV instellation, we consider the likelihood that atmospheric escape is occurring and traceable with H I Lyman~$\alpha$~(\Lya; 1216 \AA) and the meta-stable He I line (10830 \AA). TOI-1231~b's bulk density is similar to that of GJ~436~b (1.80$\pm$0.29 g cm$^{-3}$; \citealt{Maciejewski2014}), a planet well known for its vigorously escaping atmosphere \citep{Kulow2014,Ehrenreich2015}. TOI-1231's fundamental stellar properties are very similar to GJ 436's, and our PFS spectra indicate that TOI-1231's log$_{10}$ R$^{\prime}_{HK}$ = -5.06 is nearly equivalent to GJ 436's (-5.09; \citealt{BoroSaikia2018}), indicating the same level of magnetic activity. Since R$^{\prime}_{HK}$ is known to correlate well with UV emission \citep{Youngblood2017}, we assume GJ~436's synthetic X-ray and UV spectrum from \cite{Peacock2019} as a proxy for TOI-1231's. The integrated flux from 100-912 \AA~at 0.13 au is 172 erg cm$^{-2}$ s$^{-1}$, but could be as low as $\sim$53 erg cm$^{-2}$ s$^{-1}$ according to the estimates from the MUSCLES Treasury Survey \citep{France2016,Youngblood2016,Loyd2016} based on \cite{Linsky2014}. In the energy-limited approximation \citep{Salz2016}, the mass loss rate scales inversely with the planet’s bulk density and inversely with the square of the orbital distance. Thus, under this approximation we expect a mass loss rate about 14 times lower for TOI-1231~b than for GJ 436~b.

\begin{figure}
\includegraphics[width=\linewidth]{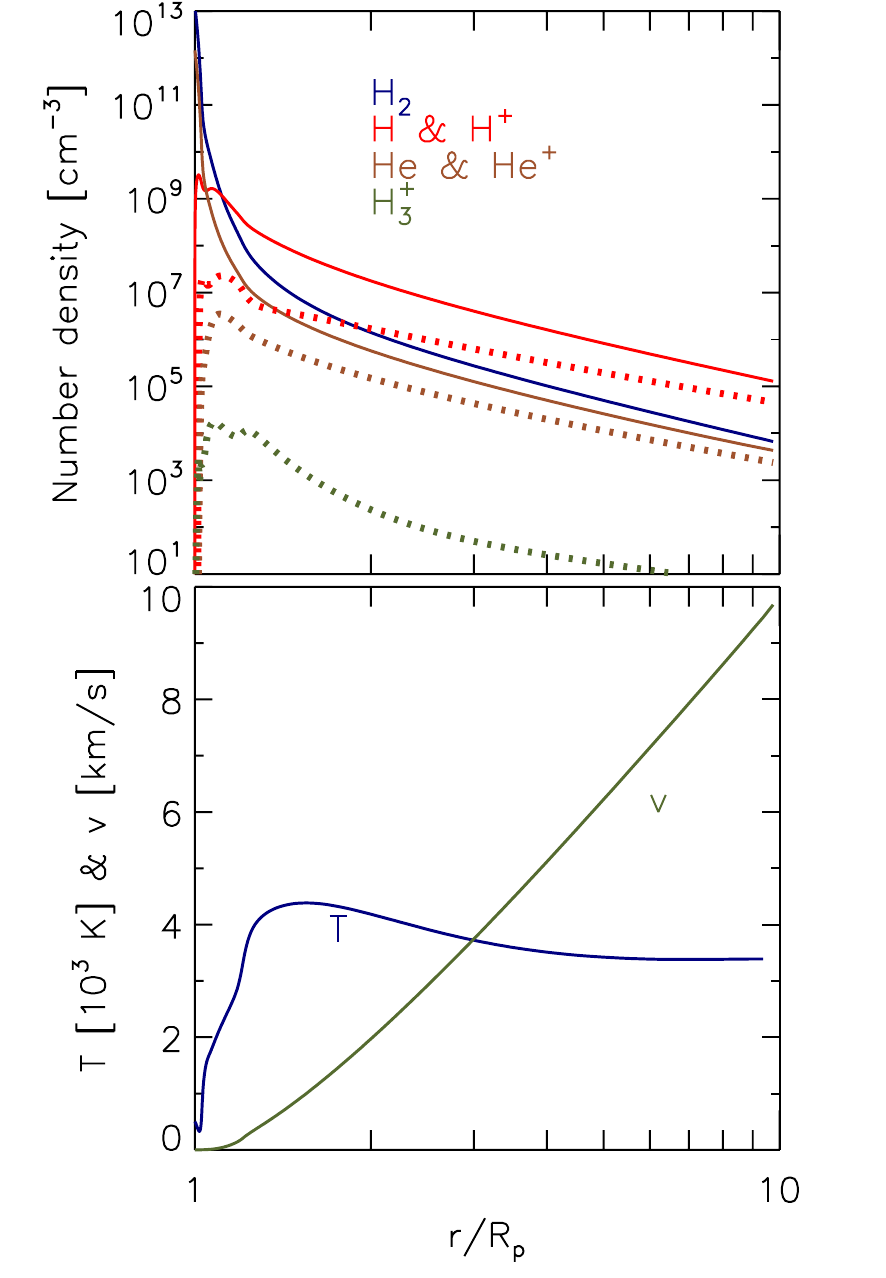}
\caption{Results from the simulated upper atmosphere of TOI-1231~b. Top panel: the number density of H$_2$ (blue), H and H$^{+}$ (red), He and He$^{+}$ (brown), and H$_3^+$ (green) are shown as a function of distance (r/R$_p$) from the planet’s 1-$\mu$m pressure level, that we
approximate as the planet’s optical radius (R$_p$). Neutrals are shown as solid lines and ions are shown as dotted lines. Bottom panel: The
temperature (10$^3$ K; blue line) and outward bulk velocity (km s$^{-1}$; green line) are shown as a function of distance from the planet’s 1-$\mu$m pressure level.
\label{fig:TOI1231_upperatmosphere}}
\end{figure}

To better understand the properties of the escaping atmosphere of TOI-1231~b, we constructed a 1D model of this planet's upper atmosphere, which solves the hydrodynamics equations for an escaping atmosphere and considers photochemistry at pressures $\lesssim$1 $\mu$bar, and radial distances from the planet center $r/R_{P}$ = 1--10. We assume the planet's bulk composition is dominated by H$_2$ and He, and set volume mixing ratios at the 1 $\mu$bar boundary of 0.9 for H$_2$ and 0.1 for He. The model is one-dimensional and spherically-symmetric, appropriate to the sub-stellar line. See \cite{GarciaMunoz2020} and references within for more details about the method. The model simulations shown in Figure~\ref{fig:TOI1231_upperatmosphere} are specific to the stellar spectrum from \cite{Peacock2019} and assume supersonic conditions at r/R$_p\sim$~10.  We take the stellar spectrum in its original format\footnote{http://archive.stsci.edu/hlsp/hazmat}, correcting only for orbital distance. A one-dimensional approach is expected to give a good representation of the flow within a few planetary radii from the surface, but cannot capture the shape of the flow after its interaction with the stellar wind. In any case, our predictions provide helpful insight to understand the prevalent forms of hydrogen, and the range of temperatures and velocities expected in the vicinity of the planet.

Figure~\ref{fig:TOI1231_upperatmosphere} shows that H$_2$ remains abundant up to very high altitudes, and that the transition from H to H$^+$ also occurs at high altitude, a condition favorable for \Lya~transit spectroscopy. Unlike for typical hot Jupiters, H$_3^+$ remains relatively abundant over an extended column and contributes to cooling of the atmosphere and to reducing the mass loss rate. The model predicts that the planet is losing 2.3$\times$10$^9$ g\,s$^{-1}$ (integrated over a solid angle 4$\pi$). 

We use the H I profile predicted by the model to estimate the \Lya~transit depth attributable to hydrogen escaping the planet and before interacting with the stellar wind. This `cold' component is typically hidden by ISM absorption and has so far remained undetected. The \Lya~absorption reported in other systems including GJ 436~b (e.g., \citealt{Ehrenreich2015}) is attributed to a `hot' component that results from charge exchange of the hydrogen atoms from the planet and the stellar wind protons \citep{Khodachenko2019}. 

\begin{figure}
\includegraphics[width=\linewidth]{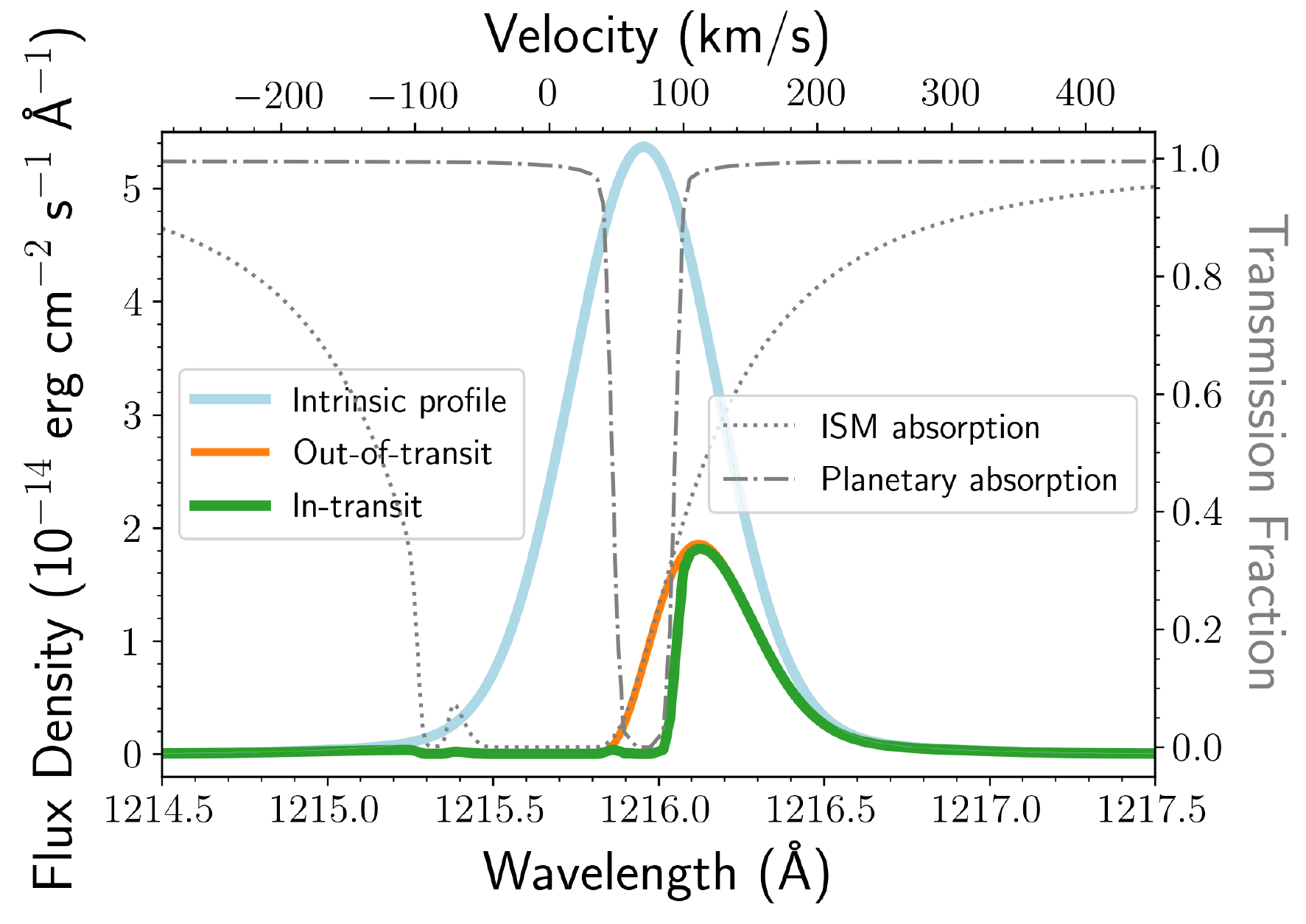}
\caption{The flux densities of TOI-1231 at Earth without (blue) and with (orange) ISM attenuation and with ISM and in-transit planetary absorption (green). TOI-1231's intrinsic \Lya~profile is scaled from GJ 436's and shifted to +70.5\,\kms, and the ISM attenuation is conservatively assumed to be log\,N(HI) = 18.6, centered at -4\,\kms\, \citep{Redfield2008}. The planetary absorption during transit is centered at 0 \kms\, in the stellar rest frame. The dotted and dash-dotted grey lines represent the ISM and in-transit planetary absorption, respectively (right axis). The transmission fraction ($e^{-\tau}$) is unity where all photons are transmitted and zero where all photons are absorbed.
\label{fig:LyA}}
\end{figure}

Despite the smaller expected escape rate with respect to other planets, TOI-1231~b's exosphere may be observable during a H I \Lya~(1215.67 \AA) transit. The host star's radial velocity (+70.5 \kms) Doppler shifts the entire system partially out of the bulk of ISM's \ion{H}{1} attenuation region, allowing access to the core of the \Lya~line and therefore to the `cold' component.

In order to assess the utility of a \Lya~transit with HST for studying this planet, we estimate the profiles of the stellar \Lya~emission, ISM attenuation, and planetary absorption (Figure \ref{fig:LyA}). We use the reconstructed Ly$\alpha$~profile of GJ 436 \citep{Youngblood2016} rescaled to match TOI-1231's distance. For the ISM, we assume a velocity centroid for the H I absorbers of -4.3\,\kms\, based on a kinematic model for the local ISM \citep{Redfield2008} and a conservative H I column density log$_{10}$ N(HI) = 18.6 based on measured column densities of nearby sight lines \citep{Wood2005,Youngblood2016,Waalkes2019}. Using STScI's online exposure time calculator for STIS with the G140M, 1222 central wavelength, and 52\arcsec$\times$0.2\arcsec\ slit, we find that the expected planetary absorption around the Ly$\alpha$~line core (the `cold' component) could be detected at high confidence in a single transit.

We have not modeled the expected transit signature for metastable He I, but note the potential for exploring this planet's upper atmosphere with He I transits. However, as noted above TOI-1231's XUV spectrum, which is responsible for both driving escape and populating the metastable He I state, is likely very similar to GJ 436's, and GJ 436~b's vigorously escaping atmosphere was not detected in He I \citep{Nortmann2018}. TOI-1231 b's larger orbit (a = 0.1266 AU) further lowers its likelihood to be a promising He I target, as the furthest planet with a confirmed He I detection thus far is WASP 107 b which has a semi-major axis of only 0.055 au and orbits a more favorable (to He I excitation) K6V star \citep{Spake2018}. See \cite{Oklopcic2018} and \cite{Oklopcic2019} for more details on the prospects tracing atmospheric escape with metastable He I.

\section{Conclusions}

We reported the \TESS\ discovery and confirmation (using several ground-based facilities) of TOI 1231~b, a temperate, Neptune-sized planet orbiting a nearby (27.6 pc) M dwarf star. The mass and radius of \thisplanetb\ were measured to be \plmassunc\, and \plradunc, respectively. By virtue of its volatile rich atmosphere, long transit duration and small host star, TOI 1231~b appears to be one of the most promising small exoplanets for transmission spectroscopy with HST and \jwst\ detected by the \TESS\ mission thus far. It represents a rare and valuable addition to the current sample of just one other low-density Neptune-sized or smaller planet with an equilibrium temperature in the 250-350 K range and a transmission spectrum (K2-18~b). Moreover, its high systemic radial velocity makes it a particularly attractive target for atmospheric escape observations via the H I Lyman~$\alpha$~, and possibly the meta-stable He I line.

This planet also serves as excellent motivation for follow up efforts focused on \TESS\ single transit events \citep{Villanueva:2019}, which is how TOI-1231~b would have presented itself if the host star was only observed for a single TESS sector. 


\clearpage
\startlongtable
\begin{deluxetable*}{llccc}
\tabletypesize{\small}
\tablecaption{Median values and 68\% confidence interval for EXOFASTv2 results on TOI-1231.\\
Notes from \citet{Eastman2019}: The optimal conjunction time ($T_{0}$)  is the time of conjunction that minimizes the covariance with the planet's period and therefore has the smallest uncertainty. The equilibrium temperature of the planet ($T_{eq}$) is calculated using Equation 1 of \citet{HansenBarman2007} and assumes no albedo and perfect heat redistribution. The tidal circularization timescale ($\tau_{\rm circ}$) is calculated using Equation 3 from \citet{AdamsLaughlin2006} and assumes Q = 10$^{6}$. The 3.6$\mu$m and 4.6$\mu$m secondary occultation depths use a black-body approximation of the stellar flux, $F_{\star}$, at T$_{\rm{eff}}$ and of the planetary flux, $F_{p}$, at T$_{\rm{eq}}$ and are calculated using $\delta_{S,\lambda}= \frac{(R_{p}/R_{\star})^{2}}{(R_{p}/R_{\star})^{2} + (F_{\star}/F_{p})}$.}
\tablehead{\colhead{~~~Parameter} & \colhead{Units} & \multicolumn{3}{c}{Values}}
\startdata
\smallskip\\\multicolumn{2}{l}{EXOFASTv2 Gaussian priors:}&\smallskip\\
~~~~$M_*$\dotfill &Stellar mass (\msun) \dotfill &$0.485\pm0.024$\\
~~~~$T_{\rm eff}$\dotfill &Effective Temperature (K) \dotfill &$3562\pm101$\\
~~~~$[{\rm Fe/H}]$\dotfill &Metallicity (dex) \dotfill &$0.05\pm0.08$\\
~~~~$\varpi$\dotfill &Parallax (mas)\dotfill &$36.3726\pm0.0163$\\
~~~~$A_V$\dotfill &V-band extinction (mag)\dotfill &$0.0031\pm0.0465$\\
\smallskip\\\multicolumn{2}{l}{Stellar Parameters:}&\smallskip\\
~~~~$M_*$\dotfill &Mass (\msun)\dotfill &$0.485\pm0.024$\\
~~~~$R_*$\dotfill &Radius (\rsun)\dotfill &$0.476^{+0.015}_{-0.014}$\\
~~~~$R_{*,SED}$\dotfill &Radius$^{1}$ (\rsun)\dotfill &$0.4766\pm0.0084$\\
~~~~$L_*$\dotfill &Luminosity (\lsun)\dotfill &$0.0326\pm0.0010$\\
~~~~$F_{Bol}$\dotfill &Bolometric Flux (10$^{-9}$ erg s$^{-1}$ cm$^{-2}$)\dotfill &$1.381^{+0.043}_{-0.044}$\\
~~~~$\rho_*$\dotfill &Density (g cm$^{-3}$)\dotfill &$6.31^{+0.68}_{-0.64}$\\
~~~~$\log{g}$\dotfill &Surface gravity (log(cm s$^{-2}$))\dotfill &$4.767^{+0.033}_{-0.035}$\\
~~~~$T_{\rm eff}$\dotfill &Effective Temperature (K)\dotfill &$3553^{+51}_{-52}$\\
~~~~$T_{\rm eff,SED}$\dotfill &Effective Temperature$^{1}$ (K)\dotfill &$3553\pm31$\\
~~~~$[{\rm Fe/H}]$\dotfill &Metallicity (dex)\dotfill &$0.041^{+0.069}_{-0.063}$\\
~~~~$A_V$\dotfill &V-band extinction (mag)\dotfill &$0.030^{+0.033}_{-0.021}$\\
~~~~$\sigma_{SED}$\dotfill &SED photometry error scaling \dotfill &$2.09^{+0.52}_{-0.37}$\\
~~~~$\varpi$\dotfill &Parallax (mas)\dotfill &$36.373\pm0.016$\\
~~~~$d$\dotfill &Distance (pc)\dotfill &$27.493\pm0.012$\\
\smallskip\\\multicolumn{2}{l}{Planetary Parameters:}&b\smallskip\\
~~~~$P$\dotfill &Period (days)\dotfill &$24.245586^{+0.000064}_{-0.000066}$\\
~~~~$R_P$\dotfill &Radius (\re)\dotfill &$3.65^{+0.16}_{-0.15}$\\
~~~~$M_P$\dotfill &Mass (\me)\dotfill &$15.4\pm3.3$\\
~~~~$T_C$\dotfill &Time of conjunction (\bjdtdb)\dotfill &$2458563.88838^{+0.00057}_{-0.00058}$\\
~~~~$T_T$\dotfill &Time of minimum projected separation (\bjdtdb)\dotfill &$2458563.88839^{+0.00057}_{-0.00058}$\\
~~~~$T_0$\dotfill &Optimal conjunction Time (\bjdtdb)\dotfill &$2458685.11630^{+0.00048}_{-0.00049}$\\
~~~~$a$\dotfill &Semi-major axis (AU)\dotfill &$0.1288^{+0.0021}_{-0.0022}$\\
~~~~$i$\dotfill &Inclination (Degrees)\dotfill &$89.73\pm0.18$\\
~~~~$e$\dotfill &Eccentricity \dotfill &$0.087^{+0.12}_{-0.061}$\\
~~~~$\omega_*$\dotfill &Argument of Periastron (Degrees)\dotfill &$176^{+81}_{-90}$\\
~~~~$T_{eq}$\dotfill &Equilibrium temperature (K)\dotfill &$329.6^{+3.8}_{-3.7}$\\
~~~~$\tau_{\rm circ}$\dotfill &Tidal circularization timescale (Gyr)\dotfill &$20600^{+9500}_{-8600}$\\
~~~~$K$\dotfill &RV semi-amplitude (\ms)\dotfill &$5.6\pm1.2$\\
~~~~$R_P/R_*$\dotfill &Radius of planet in stellar radii \dotfill &$0.0701^{+0.0019}_{-0.0017}$\\
~~~~$a/R_*$\dotfill &Semi-major axis in stellar radii \dotfill &$58.1\pm2.0$\\
~~~~$\delta$\dotfill &Transit depth (fraction)\dotfill &$0.00492^{+0.00027}_{-0.00024}$\\
~~~~$Depth$\dotfill &Flux decrement at mid transit \dotfill &$0.00492^{+0.00027}_{-0.00024}$\\
~~~~$\tau$\dotfill &Ingress/egress transit duration (days)\dotfill &$0.00946^{+0.0018}_{-0.00066}$\\
~~~~$T_{14}$\dotfill &Total transit duration (days)\dotfill &$0.1350^{+0.0017}_{-0.0014}$\\
~~~~$T_{FWHM}$\dotfill &FWHM transit duration (days)\dotfill &$0.1251\pm0.0013$\\
~~~~$b$\dotfill &Transit Impact parameter \dotfill &$0.27^{+0.19}_{-0.18}$\\
~~~~$b_S$\dotfill &Eclipse impact parameter \dotfill &$0.27^{+0.14}_{-0.18}$\\
~~~~$\tau_S$\dotfill &Ingress/egress eclipse duration (days)\dotfill &$0.00961^{+0.00085}_{-0.00077}$\\
~~~~$T_{S,14}$\dotfill &Total eclipse duration (days)\dotfill &$0.136^{+0.012}_{-0.013}$\\
~~~~$T_{S,FWHM}$\dotfill &FWHM eclipse duration (days)\dotfill &$0.126^{+0.012}_{-0.013}$\\
~~~~$\delta_{S,2.5\mu m}$\dotfill &Blackbody eclipse depth at 2.5$\mu$m (ppm)\dotfill &$0.000522^{+0.00012}_{-0.000097}$\\
~~~~$\delta_{S,5.0\mu m}$\dotfill &Blackbody eclipse depth at 5.0$\mu$m (ppm)\dotfill &$0.99^{+0.13}_{-0.11}$\\
~~~~$\delta_{S,7.5\mu m}$\dotfill &Blackbody eclipse depth at 7.5$\mu$m (ppm)\dotfill &$10.49^{+1.0}_{-0.86}$\\
~~~~$\rho_P$\dotfill &Density (g cm$^{-3}$)\dotfill &$1.74^{+0.47}_{-0.42}$\\
~~~~$logg_P$\dotfill &Surface gravity \dotfill &$3.054^{+0.094}_{-0.11}$\\
~~~~$\Theta$\dotfill &Safronov Number \dotfill &$0.079\pm0.017$\\
~~~~$\fave$\dotfill &Incident Flux (10$^{9}$ erg s$^{-1}$ cm$^{-2}$)\dotfill &$0.00263^{+0.00013}_{-0.00014}$\\
~~~~$T_P$\dotfill &Time of Periastron (\bjdtdb)\dotfill &$2458543.7^{+4.8}_{-6.3}$\\
~~~~$T_S$\dotfill &Time of eclipse (\bjdtdb)\dotfill &$2458575.45^{+0.95}_{-2.3}$\\
~~~~$T_A$\dotfill &Time of Ascending Node (\bjdtdb)\dotfill &$2458557.57^{+0.57}_{-1.4}$\\
~~~~$T_D$\dotfill &Time of Descending Node (\bjdtdb)\dotfill &$2458569.62^{+0.63}_{-1.1}$\\
~~~~$V_c/V_e$\dotfill & \dotfill &$0.993^{+0.062}_{-0.048}$\\
~~~~$e\cos{\omega_*}$\dotfill & \dotfill &$-0.036^{+0.061}_{-0.15}$\\
~~~~$e\sin{\omega_*}$\dotfill & \dotfill &$0.002^{+0.046}_{-0.066}$\\
~~~~$M_P\sin i$\dotfill &Minimum mass (\me)\dotfill &$15.4\pm3.3$\\
~~~~$M_P/M_*$\dotfill &Mass ratio \dotfill &$0.000096^{+0.000021}_{-0.000020}$\\
~~~~$d/R_*$\dotfill &Separation at mid transit \dotfill &$57.5^{+4.3}_{-4.4}$\\
~~~~$P_T$\dotfill &A priori non-grazing transit prob \dotfill &$0.0162^{+0.0014}_{-0.0011}$\\
~~~~$P_{T,G}$\dotfill &A priori transit prob \dotfill &$0.0186^{+0.0016}_{-0.0013}$\\
~~~~$P_S$\dotfill &A priori non-grazing eclipse prob \dotfill &$0.01607^{+0.0016}_{-0.00069}$\\
~~~~$P_{S,G}$\dotfill &A priori eclipse prob \dotfill &$0.01849^{+0.0019}_{-0.00080}$\\
\smallskip\\\multicolumn{2}{l}{Wavelength Parameters:}&R&z'&TESS\smallskip\\
~~~~$u_{1}$\dotfill &linear limb-darkening coeff \dotfill &$0.31^{+0.33}_{-0.22}$&$0.20^{+0.25}_{-0.15}$&$0.31\pm0.19$\\
~~~~$u_{2}$\dotfill &quadratic limb-darkening coeff \dotfill &$0.36^{+0.33}_{-0.44}$&$0.22^{+0.34}_{-0.27}$&$0.07^{+0.29}_{-0.23}$\\
~~~~$A_D$\dotfill &Dilution from neighboring stars \dotfill &--&--&$-0.001^{+0.046}_{-0.049}$\\
\smallskip\\\multicolumn{2}{l}{Telescope Parameters:}&PFS velocities\smallskip\\
~~~~$\gamma_{\rm rel}$\dotfill &Relative RV Offset (m/s)\dotfill &$0.11\pm0.91$\\
~~~~$\sigma_J$\dotfill &RV Jitter (m/s)\dotfill &$3.03^{+1.0}_{-0.72}$\\
~~~~$\sigma_J^2$\dotfill &RV Jitter Variance \dotfill &$9.2^{+7.2}_{-3.8}$\\
\enddata
\label{tab:EXOFASTv2}
\tablenotetext{1}{This value ignores the systematic error and is for reference only}
\end{deluxetable*}

\acknowledgements

This paper includes data collected by the \TESS\ mission. Funding for the \TESS\ mission is provided by NASA's Science Mission directorate. We acknowledge the use of public \TESS\ Alert data from pipelines at the \TESS\ Science Office and the \TESS\ Science Operations Center. 
This paper includes data gathered with the 6.5 meter Magellan Telescopes located at Las Campanas Observatory, Chile. This work makes use of observations from the LCOGT network and from the ASTEP telescope. ASTEP benefited from the support of the French and Italian polar agencies IPEV and PNRA in the framework of the Concordia station program. 
This work has made use of data from the European Space Agency (ESA) mission Gaia (https://www.cosmos.esa.int/gaia), processed by the Gaia Data Processing and Analysis Consortium (DPAC, https://www.cosmos.esa.int/web/gaia/dpac/consortium). Funding for the DPAC has been provided by national institutions, in particular the institutions participating in the Gaia Multilateral Agreement. 
Some of the observations in the paper made use of the High-Resolution Imaging instrument Zorro. Zorro was funded by the NASA Exoplanet Exploration Program and built at the NASA Ames Research Center by Steve B. Howell, Nic Scott, Elliott P. Horch, and Emmett Quigley. Zorro was mounted on the Gemini South telescope of the international Gemini Observatory, a program of NSF’s OIR Lab, which is managed by the Association of Universities for Research in Astronomy (AURA) under a cooperative agreement with the National Science Foundation on behalf of the Gemini partnership: the National Science Foundation (United States), National Research Council (Canada), Agencia Nacional de Investigación y Desarrollo (Chile), Ministerio de Ciencia, Tecnología e Innovación (Argentina), Ministério da Ciência, Tecnologia, Inovações e Comunicações (Brazil), and Korea Astronomy and Space Science Institute (Republic of Korea).
Resources supporting this work were provided by the NASA High-End Computing (HEC) Program through the NASA Advanced Supercomputing (NAS) Division at Ames Research Center for the production of the SPOC data products. 
Some of the data presented in this paper were obtained from the Mikulski Archive for Space Telescopes (MAST). 
Support for MAST for non-HST data is provided by the NASA Office of Space Science via grant NNX13AC07G and by other grants and contracts. 
This research has made use of the NASA Exoplanet Archive, which is operated by the California Institute of Technology, under contract with the National Aeronautics and Space Administration under the Exoplanet Exploration Program. 
This research has made use of NASA's Astrophysics Data System. 
This research has also made use of the Exoplanet Follow-up Observation Program website, which is operated by the California Institute of Technology, under contract with the National Aeronautics and Space Administration under the Exoplanet Exploration Program. 
This research made use of Astropy, a community-developed core Python package for Astronomy \citep{Astropy2013}. 
Part of this research was carried out at the Jet Propulsion Laboratory, California Institute of Technology, under a contract with the National Aeronautics and Space Administration (NASA).
D. D. acknowledges support from the TESS Guest Investigator Program grant 80NSSC19K1727 and NASA Exoplanet Research Program grant 18-2XRP18\_2-0136.
TD acknowledges support from MIT's Kavli Institute as a Kavli postdoctoral fellow.
E.M. acknowledges support from NASA award 17-K2G06-0030.
P.M. acknowledges support from the European Research Council under the European Union's Horizon 2020 research and innovation program under grant agreement No. 832428.
T.G., A.A., L.A., D.M., F.-X.S. acknowledge support from Idex UCAJEDI (ANR-15-IDEX-01).
D.J.S. acknowledges funding support from the Eberly Research Fellowship from The Pennsylvania State University Eberly College of Science. The Center for Exoplanets and Habitable Worlds is supported by the Pennsylvania State University, the Eberly College of Science, and the Pennsylvania Space Grant Consortium.
This research received funding from the European Research Council (ERC) under the European Union's Horizon 2020 research and innovation programme (grant agreement n$^\circ$ 803193/BEBOP), and from the Science and Technology Facilities Council (STFC; grant n$^\circ$ ST/S00193X/1).
This publication makes use of VOSA, developed under the Spanish Virtual Observatory project supported by the Spanish MINECO through grant AyA2017-84089. VOSA has been partially updated by using funding from the European Union's Horizon 2020 Research and Innovation Programme, under Grant Agreement nº 776403 (EXOPLANETS-A).

\facilities{TESS, Magellan:Clay (Planet Finder Spectrograph), Gemini-South (Zorro), SOAR, LCOGT, ASTEP} 

\software{AstroImageJ \citep{Collins:2017}, TAPIR \citep{Jensen:2013}, EXOFASTv2 \citep{Eastman2019}, petitCODE \citep{mollierevanboekel2017}, PyMultiNest \citep{Buchner2014}, MultiNest \citep{ferozhobson2009}, Astropy \citep{Astropy2013}}

\bibliographystyle{apj}



\end{document}